\documentclass[apj, iop]{emulateapj}
\usepackage{amsmath}
\usepackage{amssymb}
\usepackage{graphicx,graphics}
\usepackage[english]{babel}
\usepackage{color}

\def\PGRAPE{$\varphi$GRAPE }
\begin{document}

\title{Supermassive Black Holes in Galactic Nuclei with Tidal Disruption of Stars: Paper II - Axisymmetric Nuclei}

\author{
       Shiyan Zhong   \altaffilmark{1},
       Peter Berczik  \altaffilmark{1,2,3},
       Rainer Spurzem \altaffilmark{1,2,4,5}
       }

\altaffiltext{1}{National Astronomical Observatories of China and Key Lab for Computational Astrophysics, Chinese Academy of Sciences, 20A Datun Rd., Chaoyang District, 100012, Beijing, China}
\altaffiltext{2}{Astronomisches Rechen-Institut, Zentrum f\"ur Astronomie, University of Heidelberg, M\"onchhofstrasse 12-14, 69120, Heidelberg, Germany}
\altaffiltext{3}{Main Astronomical Observatory, National Academy of Sciences of Ukraine, 27 Akademika Zabolotnoho St., 03680, Kyiv, Ukraine}
\altaffiltext{4}{Kavli Institute for Astronomy and Astrophysics, Peking University, Beijing, China}
\altaffiltext{5}{Key Lab of Frontiers in Theoretical Physics, Institute of Theoretical Physics, Chinese Academy of Sciences, Beijing, 100190, P.R. China}

\shorttitle{Tidal Disruption by SMBH}
\shortauthors{Zhong, Berczik, Spurzem}

%%%%%%%%%%%%%%%%%%%%%%%%%%%%%%%%%%%%%%%%%%      ABSTRACT

\begin{abstract}

Tidal Disruption of stars by supermassive central black holes from dense rotating star clusters is modelled
by high-accuracy direct $N$-body simulation. As in a previous paper on spherical star clusters we study the
time evolution of the stellar tidal disruption rate and the origin of tidally disrupted stars, now according to
several classes of orbits which only occur in axisymmetric systems (short axis tube and saucer).
Compared with that in spherical systems,
we found a higher TD rate in axisymmetric systems. The enhancement can be explained by an enlarged
loss-cone in phase space which is raised from the fact that total angular momentum $\bf J$ is not conserved.
As in the case of spherical systems, the distribution of the last apocenter distance of tidally accreted
stars peaks at the classical critical radius. However, the angular distribution of the origin of the
accreted stars reveals interesting features. Inside the influence radius of the supermassive black hole
the angular distribution of disrupted stars has a conspicuous bimodal structure with a local
minimum near the equatorial plane. Outside
the influence radius this dependence is weak. We show that the bimodal structure of orbital
parameters can be explained by the presence of two
families of regular orbits, namely short axis tube and saucer orbits. Also the consequences of our
results for the loss cone in axisymmetric galactic nuclei are presented.

\end{abstract}


%%%%%%%%%%%%%%%%%%%%%%%%%%%%%%%%%%%%%%%%%%      Introduction
\section{Introduction}

A large fraction of galaxies show evidence of supermassive black holes (henceforth SMBH) residing in their center.
They are typically embedded in nuclear star clusters (NSC); if resolution allows to observe the NSCs,
they are among the densest clusters known.
Their size is similar to galactic globular clusters, but they are much heavier and brighter
\citep{BLV2002,BSM2004}. In massive galaxies NSCs may not be significant or even do not exist, however,
the SMBHs are still surrounded by enormous number of stars. SMBH residing in these NSCs will tidally
disrupt stars that come close to its tidal radius and eventually accrete the gaseous debris, which can
light up the central SMBH for a period of time \citep{Rees1988,EK1989}. This kind of event is a useful
tool to examine the relativistic physics near SMBH since the disruption occurs at a place very close
to the BH's Schwarzschild radius. Also it can help us to investigate SMBH in non-active galactic
center. Although tidal disruption of stars has been proposed for almost half a century, only until
last decade do people realize the importance of such events, after the discovery of a dozens of tidal
disruption candidates \citep{Komossa2002,KM2008}. \citet{LLK2014} discovered a candidate of binary
SMBH system by analyzing the break in the light curve of TD event, demonstrate it as a promising tool
for searching hidden SMBH binaries in quiescent galactic center. In order to compute the tidal
disruption event rate, many theoretic works have been
done in the past few decades \citep{FR1976,LS1977,MT1999,WM2004}. The core of the story is loss cone
theory, which was first established in the case of spherical symmetric systems.

Stars with orbital pericenter smaller than the tidal radius $r_t$ are defined to be inside the loss cone,
with $r_t$ be expressed by

\begin{equation}
r_t = \alpha r_\star \frac{3}{5-n} \Bigl(\frac{M_{\rm bh}}{ m_\star} \Bigr)^{\frac{1}{3}}\ ,
\end{equation}

where $r_\star$, $m_\star$ are the radius and mass of a star, $n$ is its polytropic index
(assuming the stellar structure can be approximated by a polytropic sphere) and
$\alpha$ is a free parameter used by us for scaling.
Stars with angular momentum $J < J_{lc} \approx
\sqrt{2 G M_{\bullet} r_t}$ are inside the loss cone. Typically, loss cone
stars are consumed in dynamical time scales. If no new star is
supplied to loss cone, there will be no more tidal disruption
event. Based on the status of loss cone, it can be divided into two
regime, namely empty and full loss cone. Due to the short ``lifetime"
of the loss cone stars, loss cone will become empty quickly. Refilling
of loss cone happens in relaxation timescales and is often referred to as
diffusion process in angular momentum space. Thus in empty loss cone
regime it is the refilling rate which controls the disruption rate.
Note that throughout this paper, and like in most if not all of the cited
papers on tidal accretion of stars onto SMBH, we assume that a star is
disrupted completely at $r_t$ and all its mass, energy and angular momentum
absorbed momentarily by the SMBH. We know that this is not realistic, and more
detailed numerical models of the process of disruption, possible disk formation
and accretion show that only fractions of the material are absorbed into the
SMBH after a number of orbits \citep{GRR2013,HSL2013}. However, the assumption that the process is fast is reasonable compared to the orbital time scales of stars further out in the cluster.

In a previous paper (\citet{ZBS2014}, henceforth Paper I) we have
shown that the classical loss cone approximation, for a spherically symmetric
system in the diffusive empty loss cone regime,
can be well reproduced by large direct $N$-body models with tidal accretion of stars
onto SMBH. Now we are focusing on the generalization to axisymmetric galactic nuclei
and compare our new results in an otherwise very similar study to those of Paper I.

Tidal disruption of stars is one possible way of growth for SMBH, especially
in quiescent galactic nuclei. Since most models assumed spherical stellar
clusters, SMBH growth rates by tidal disruption are very low, limited by the
very long relaxation time to refill the loss cone, and the contribution of
the process to the overall growth of SMBH is considered as relatively insignificant.
However, the stellar distribution in real galactic
nuclei might not be spherically symmetric. Many galactic nuclei show evidence
of rotation in their centers, even very close to the SMBH \citep{MMH1995,NM1995,GHB1995}.

According to the current standard model of structure formation massive galaxies
have undergone quite significant mergers (in number and mass ratio). Numerical
models of the merging process of galaxies show that the merger remnant shows
rotation, axial symmetry or even triaxiality in the central regions \citep{KJM2011,PBB2011,GM2012,BEB2013}.

In the center of our own Milky Way the NSC can be observed in unparalleled
high resolution \citep{FNS2014,SFK2014}. It consists of
$1.4\times10^{7} M_{\odot}$ within its effective radius ($4.2$
pc); kinematic data indicate that it possesses bulk rotation
\citep{FNS2014}. The formation mechanism of NSCs is still under
debate. There are two scenarios, \textit{in situ}
formation \citep{Milosav2004} and a sinking scenario (globular cluster sink to the center
and merge) \citep{TOS1975,LTF2001}. NSCs in a sample of nearby
galaxies observed by \citet{SDH2006,SBB2008} show that these objects
are non-spherical and even contain multi-component (younger disk plus older spherical component),which favor the
\textit{in situ} scenario. However, \citet{ACM2012} have performed a
series of $N$-body simulations to study the formation of NSCs, which
support the sinking scenario. The model NSCs formed in their simulations
by merging between infalling globular clusters initially have mildly
triaxial shape. After the final infall, the shape of the NSC will
gradually become axisymmetric in following dynamical evolution.

Despite the debate between different formation scenarios, we think that
it is quite likely
that NSC are non-spherical. This provides a good motivation
to study the tidal disruption rate in axisymmetric (and triaxial) clusters. Some
works have already been done but the mission is not
over. \citet{FS2010} used 2D Fokker-Plank model \citep{ES1999,KEL2002}
to study rotating dense stellar clusters with BHs and cross checked
with $N$-body models \citep{FPB2012}. Both works find that BH embedded
in rotating model have higher tidal disruption rate (hereafter TDR) compare to spherical models. BH mass
at the end of simulation is roughly 20\% higher in rotating case. They
find an excess of accreted prograde rotating stars which are
originated mainly outside the influence radius $r_h$ and call for a
further investigation of the roles of stars with non-conserved $J_x,
J_y$ angular momentum. As shown by the works of \citet{MP2004}, in
non-spherical systems chaotic orbits (existing in regions outside
$r_h$) can keep the loss cone full for sufficient long time, thus
tidal disruption can contribute a lot of mass within Hubble time and
could play an important role in the BH growth across cosmic time.

On the other hand, the loss cone itself might be enlarged as pointed
out by \citet{MT1999}, due to the fact that angular momentum $\bf{J}$
is not conserved in axisymmetric potential. \citet{VM2013}
confirmed this picture in a detailed analysis of the loss cone problem
in axisymmetric galactic nuclei. They analyzed the depletion and refilling
of loss cone orbits and found that tidal disruption rates could be increased
by a moderate factor due to axisymmetry as compared to spherical symmetry.
In their work chaotic orbits with low angular momentum, which can reach
just outside the influence radius at apocenter, but also get close to
the central SMBH at pericenter, cause some difficulty in comparison with
Fokker-Planck models, as was already found by \citet{MVN1993} (Note
that the last author of this paper is the same person than the last author
of \citet{MVN1993}, there was a mistake in re-translating the name from
Russian language).

In this work we follow an experimental numerical approach to the problem,
following Paper I for the case of spherically symmetric systems. We treat
particle number and tidal radius as free parameters and analyze the tidal
accretion rate of the system as a function of the strength of deviation
from spherical symmetry. We
measure the shape of the loss cone in axisymmetric potential and
and characterize the characteristic orbits of stars in the loss cone.
We find
that it is indeed enlarged and can account for the higher
TDR as compared to spherically symmetric galactic nuclei.

This paper is organized as follows: we describe the model setup of the
simulation in Section 2 and present the result of TDR measurement in
Section 3. Section 4 is devoted to the measurement of loss cone shape
in axisymmetric potential and we demonstrate the enlargement of loss
cone. In Section 5, we present the result for the origin and orbital
classification of disrupted stars. In Section 6, we discuss the
potential application of our results.

%%%%%%%%%%%%%%%%%%%%%%%%%%%%%%%%%%%%%%%%%%      Model
\section{$N$-body model}

We adopt the standard $N$-body unit definitions from \citet{HM1986}, namely $G$ = $M$ = 1
and $E$ = $-1/4$, where $G$ is the gravitational constant, $M$ is the
total mass of the model cluster and $E$ is the total energy. In our
$N$-body models we assume that all the particles have the same mass,
so $m=1/N$, where $m$ is the particle mass and $N$ is the total particle number.
To preserve the scale invariance of our $N$-body simulations we fix the
initial black hole mass relative to the total mass of the star cluster
(0.01) and use the particle number and the tidal radius $r_t$ in $N$-body units
(which is a dimensionless number) as free parameters. We have shown
in Paper I that the method of scaling to realistic parameters for $N$ and $r_t$
can be used to obtain astrophysically meaningful results from the collection of
our models. In order to support our scaling procedure we even do not change
the tidal radius during the simulations -
since the BH mass changes within one order of magnitude only
during the simulation, relative changes in tidal radius are small
(notice that $r_{t}\propto(M_{\bullet})^{1/3}$).

The initial distribution of particles follows a generalized King model with rotation.
The distribution function is \citep{ES1999,EGF2007}

\begin{equation}\label{Eq_King}
    f(E,J_z) = C\cdot[\rm exp(-\frac{E}{\sigma_{K}^2}) - 1]\cdot\rm exp(-\frac{\Omega_0 J_z}{\sigma_{K}^2}) ,
\end{equation}

where $\sigma_{K}$ is the King velocity dispersion and $\Omega_0$ is a
characteristic angular velocity. Since we are considering an isolated
system, the $\Phi_t$ is set to 0. This rotating King model has two
dimensionless parameters: $W_0$ and $\omega_0$. The King parameter
$W_0 = -\Phi_0/\sigma_{K}^2$, where $\Phi_0$ is the central potential,
controls the degree of central concentration. And the rotation
parameter $\omega_0 = \sqrt{9/(4 \pi G \rho_0)}\cdot\Omega_0$, where
$\rho_0$ is the central density, controls the degree of rotation.
$\omega_0 = 0$ will reduce the model to a usual non-rotating
spherically symmetric King model.

We limit our current study to only one concentration parameter
$W_0 = 6$ and two rotation parameters $\omega_0=0.3, 0.6$; the
density profile of King model with this concentration is similar to that
of the Plummer model used in Paper I, so it is possible to compare
with the previous results and focus on the effects of rotation and
axial symmetry only. The rotation is moderate (cf. e.g. \citet{ES1999})
and resembles that of Milky Way globular clusters.

For completeness we also employ non-rotating King model with $W_0 = 6$ and
$\omega_0 = 0.0$, which is used as a fiducial model and also a bridge to the results of Paper I,
confirming our claim that it indeed closely resembles the results for the Plummer model
used in Paper I (e.g. in the evolution of the TDR). In another test run we used a larger rotation with
$\omega_0 = 0.9$ - it experienced an unstable stage during which a bar formed but quickly
disappeared. This bar formation could probably be identified with the
radial orbit instability of \citet{AM1990}. We note that our standard models
with $\omega_0=0.3, 0.6$ remain fully axisymmetric during the entire simulation;
to study tidal disruption in triaxial systems with bars is beyond the scope of our current paper.

\begin{figure}[htbp] \begin{center}
  \includegraphics[width=\columnwidth]{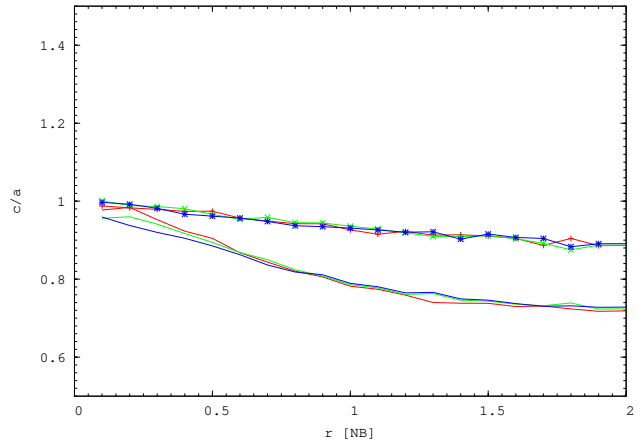}
  \end{center}
  \caption{Axial ratio for rotating models as a function of radius. For each model we show the axial ratio measured
  at different evolution stage: T=0 (red); T=500 (green); T=1000 (blue). Lines with symbols are results for $\omega_0 = 0.3$ model, lines without symbols are for $\omega_0 = 0.6$ models.}
  \label{fig_AxisRatio}
\end{figure}

Fig.~\ref{fig_AxisRatio} shows the axial ratio ($c/a$) of the model clusters
as a function of radius up to $r = 2.0$ (within which most of stars are located).
We estimate the axial ratio for both rotating models, using the moment of inertia tensor
measured in concentric shells. One can see $c/a$ is close to 1 at the
innermost part and decreases outward: $\omega_0=0.3$ model decreases slowly
to its minimum value 0.9; $\omega_0=0.6$ model decreases faster and has a minimum value 0.71.
If we measure the $c/a$ for the whole cluster, the results for the two models
are 0.9 ($\omega_0=0.3$) and 0.75 ($\omega_0=0.6$). Fig.~\ref{fig_AxisRatio}
also shows that $c/a$ is almost unchanged during long time evolution, except
for the inner part of $\omega_0=0.6$ model, which exhibits slight decrease.

In rotating systems, there is a phenomenon called
gravo-gyro instability, which is caused by the negative specific
moment of inertia \citep{IH1978,Hachisu1979,Hachisu1982}. This kind of
instability happens in long term evolution of rotating cluster which
is much longer than our integration time \citep{EGF2007}.

The model set is summarized in Table~\ref{table_model}.

%%%%%%%%%%%%%%%%%%%%%%%%%%%%%%%%%%%%%%%%%%%%%%%%%%%%
\begin{table}[htbp]
\begin{center}
\caption{Full set of our model runs.\label{table_model}}
\begin{tabular}{c|cccc}
  \tableline
  Model &   N/K    & $\omega_0$ &  $r_{t}$     &   T      \\
  \hline
  R20w00    &   64    & 0.0  &  $10^{-3}$     &  1500    \\
  R30w00    &  128    & 0.0  &  $10^{-3}$     &  1600    \\
  R21w00    &   64    & 0.0  &  $10^{-4}$     &  1500    \\
  R31w00    &  128    & 0.0  &  $10^{-4}$     &  1300    \\
  \hline
  R20w03    &   64    & 0.3  &  $10^{-3}$     &  1500    \\
  R30w03    &  128    & 0.3  &  $10^{-3}$     &  1500    \\
  R21w03    &   64    & 0.3  &  $10^{-4}$     &  2600    \\
  R31w03    &  128    & 0.3  &  $10^{-4}$     &  2000    \\
  \hline
  R20w06    &   64    & 0.6  &  $10^{-3}$     &  1500    \\
  R30w06    &  128    & 0.6  &  $10^{-3}$     &  1500    \\
  R21w06    &   64    & 0.6  &  $10^{-4}$     &  1600    \\
  R31w06    &  128    & 0.6  &  $10^{-4}$     &  2000    \\
  \tableline
\end{tabular}
\tablecomments{Column 1 : Model codename. Column 2 : Particle number in
the unit of K(=1024). Column 3 : dimensionless rotation parameter.
Column 4 : black hole's tidal radius. Column 5 : total integration time.
$r_{t}$ and $T$ are in model unit.}
\end{center}
\end{table}
%%%%%%%%%%%%%%%%%%%%%%%%%%%%%%%%%%%%%%%%%%%%%%%%%%%%

We run the simulation for more than one initial half-mass relaxation
time ($t_{rh}$), which is estimated using the same formula in Paper I
and the values can be found there as well (Table 2).

All simulations are running with the \PGRAPE code \citep{BNZ2011},
which runs with high performance (up to 350 Gflop/s per GPU) on our
GPU clusters in Beijing (NAOC/CAS). This code is a direct $N$-body
simulation package, with a high order Hermite integration scheme and
individual block time steps. A direct $N$-body code evaluates in
principle all pairwise forces between the gravitating particles, and
its computational complexity per crossing time scales asymptotically with $N^2$;
however, it is {\em not} to be confused with a simple brute force
shared time step code, due to the block time steps. We refer more
interested readers to a general discussion about $N$-body codes and
their implementation in \cite{spurzem2011a,spurzem2011b}. The present
code is well tested and already used to obtain important results in
our earlier large scale few million body simulation ~\citep{KBB2012}.

%%%%%%%%%%%%%%%%%%%%%%%%%%%%%%%%%%%%%%%%%%%%%%%%%%%%
\section{Tidal Disruption Rate (TDR)}
\subsection{Results of our work}

\begin{figure}[htbp] \begin{center}
  \includegraphics[width=\columnwidth]{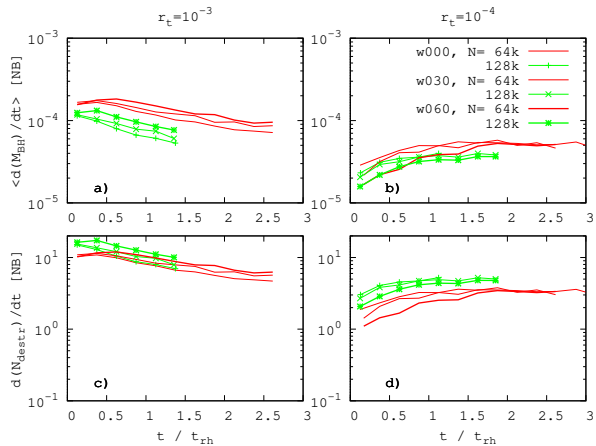}
  \end{center}
  \caption{TDR as a function of time in units of initial half mass relaxation time; averaged over intervals of $1/4 t_{rh}$. Top panels: mass accretion rate; bottom panels: particle accretion rate; left and right panels for two different tidal radii as indicated. Curves with symbols are stand for 128K models, those without symbols are stand for 64K models. Line thickness indicate different rotating parameters.}
  \label{fig_Mdot_comp-multi}
\end{figure}

In this section, we present the TDR measured in simulations with our
rotating King models and compare it with the TDR of the non-rotating model of Paper I. In
Fig. \ref{fig_Mdot_comp-multi}, we show the TDR (both in terms of
mass and particle number) as it evolves with time for two different tidal radii; in each
panel two different rotation parameters are plotted together with the
data of the non-rotating system. The time is given in units of
initial half mass relaxation time $t_{rh}$, which is convenient for
comparison of simulations with different particle numbers.
To smooth out fluctuations due to particle noise we have plotted in the
figure the TDR averaged over a time interval
(here $1/4 t_{rh}$).

The TDR with a large tidal radius (i.e. $r_t = 10^{-3}$)
initially quickly rises in the $N = 64 K$ model to its peak value,
and then decreases; for the $N = 128 K$ model the TDR
almost decreases from the beginning. The initial phase is
connected with the formation of a central density cusp in the
surrounding stellar system and with the process of transition
from initially full to empty loss cone. The BH gains mass from the accreted
stars, thus the mass ratio between stars and the BH ($\gamma := m/M_{\bullet}$)
decreases with time, and as a result the BH's random motion damps.
We have discussed in Paper I that the status of the loss-cone is
connected with the BH's Brownian motion in the sense that once the
amplitude of Brownian motion is smaller than $~10 r_t$ the system
enters the empty loss-cone regime, during
which the cusp and central density are still growing but TDR begins to
fall. In the $N = 128 K$ model, the mass ratio $\gamma$ is smaller,
so the initial loss cone depletion is very short, practically invisible
in the plots, and the subsequent evolution is determined by cusp formation and
damping of BH motion.

In the models with small tidal radius ($r_t = 10^{-4}$), there
is always an initial growth phase of TDR, followed by the
convergent approach to a stationary state.
Due to the small $r_t$ their BH growth is slow, thus
they need more time to achieve the mass required to limit their
Brownian motion.

Fig. \ref{fig_Mdot_comp-multi} also shows the TDR dependence on
rotation parameter $\omega_0$ as a new result compared to Paper I.
For large tidal radius ($r_t=10^{-3}$),
faster rotation will result in a higher TDR, note that these models
are in empty loss-cone regime.
Table \ref{table_TDR} list out the numbers
for TDR measurement. One can see $\omega_0=0.3$ model has a TDR on average
13 percent higher than $\omega_0=0.0$ model. And TDR in $\omega_0=0.6$
model is on average 35 percent higher than that in $\omega_0=0.0$ model.
BH mass of these 3 models measured at $T=1500$ are 0.131, 0.143 and 0.167. The
fractional increase of final BH mass with increasing degree of rotation is
consistent with the result of \citet{FPB2012}.
The reason for this dependence of $\omega_0$ is that in these systems
the effective loss-cone is larger than classic one in spherical system.
We will investigate such an enlarged loss-cone in more detail in the next
section. For small tidal radius ($r_t = 10^{-4}$), however,
we observe a different behavior of TDR. From beginning to about
$\sim1.5 t_{rh}$, faster rotation result in a smaller TDR!
The argument presented by \citet{MT1999} may provide some hints: if BH's
wandering time-scale is shorter than dynamical time-scale, a decrease
in TDR will happen. We note in the simulation at this early stage the BH
is quickly wandering due to its small mass and slow growth.
Furthermore, in axisymmetric systems a star's
pericenter distance changes with time (even ignoring irregular
perturbations from other stars). So when the BH comes back to the place
where it was, it may still miss the star which is supposed to be
disrupted shortly before.

\begin{table}[htbp]
\begin{center}
\caption{TDR results for $r_t = 10^{-3}$ models.\label{table_TDR}}
\begin{tabular}{c|c|cc|cc}
  \tableline
  $t/t_{rh}$ &  $\dot{N}_0$ & $\dot{N}_3$ & $\dot{N}_3/\dot{N}_0$ & $\dot{N}_6$ & $\dot{N}_6/\dot{N}_0$ \\
  \hline
  0.25   &  14.96    & 15.41  &  1.03   &  16.30  &  1.09  \\
  0.50   &  12.89    & 13.67  &  1.06   &  17.30  &  1.34  \\
  0.75   &  10.44    & 12.07  &  1.16   &  14.52  &  1.39  \\
  1.00   &   8.73    & 10.32  &  1.18   &  12.65  &  1.45  \\
  1.25   &   7.97    &  9.73  &  1.22   &  11.09  &  1.39  \\
  1.50   &   6.99    &  7.93  &  1.13   &  10.04  &  1.44  \\
  \tableline
\end{tabular}
\tablecomments{Measured TDR for models with same N (128K) and $r_t$ ($10^{-3}$) but
 different rotating parameters, at different evolution time. $\dot{N}_0$ is TDR in classic King
 model ($\omega_0=0$); $\dot{N}_3$ and $\dot{N}_6$ are results for $\omega_0=0.3$ and $\omega_0=0.6$ models.
 We also give the boost factor $\dot{N}_3/\dot{N}_0$ and $\dot{N}_6/\dot{N}_0$.}
\end{center}
\end{table}

Afterwards the system begins to enter the empty loss cone regime, and all
TDR curves converge to each other; for small tidal radius more
tidally disrupted stars originate from inside the BH influence radius,
where the system is approximately spherically symmetric. Any deviation
from spherical symmetry in our rotating models prevails near and outside
the influence radius. Convergence of TDR reflects the original results
obtained in Paper I for spherical systems.

\begin{figure}[htbp] \begin{center}
  \includegraphics[width=\columnwidth]{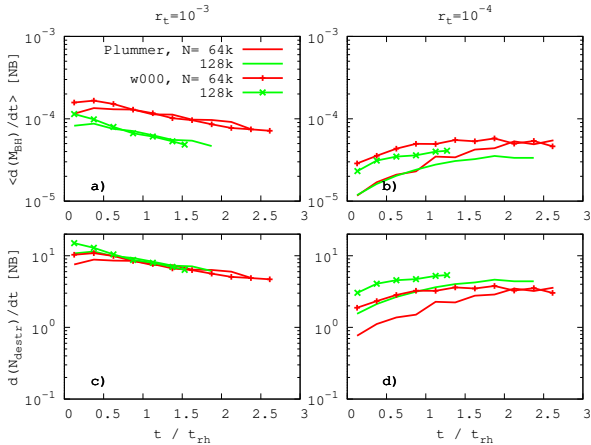}
  \end{center} \caption{$x$ axis is time in unit of initial half-mass
  relaxation time. $y$ axis for panel a) and b) is the averaged mass
  accretion rate in given time range (i.e. 1/4 $t_{rh}$); $y$ axis for
  panel c) and d) is the number accretion rate. Panel a) and c) show
  the result for $r_t = 10^{-3}$. Panel b) and d) show the result for
  $r_t = 10^{-4}$.}  \label{fig_Mdot_comp-Plum_w000} \end{figure}

Fig. \ref{fig_Mdot_comp-Plum_w000} compares the TDR of classic King
model ($W = 6, \omega_0 = 0.0$) with that in Plummer model. In $r_t =
10^{-3}$ models, except the initial higher accretion rate in King
model, the two models have similar TDR in following evolution. While
in the case of $r_t = 10^{-4}$, King model have a higher accretion
rate during most of the time, but later on they gradually come to the same
level as the Plummer model. The higher rate in King model could be
explained by the slightly higher density in the core region at
beginning. In the following evolution of $r_t = 10^{-3}$ models, the
two models form cusp similar to each other so they have roughly same
accretion rate. In the case of $r_t = 10^{-4}$, the initial accretion
rate ratio $\dot{N}_{King}/\dot{N}_{Plum}$ is higher than those in
$r_t = 10^{-3}$. BH inside King cluster growing faster and also the
growth of cusp, in the following evolution King model always have a
higher density in the cusp which in return gives a higher accretion
rate. Only after the BH gain enough mass and become a ``static"
object, the accretion rate slowly reaches a maximum and begins to drop
afterward.

Up to this point, all results were presented in model units ($N$-body units).
As in Paper I (see Sect. 5 and Appendix therein) we will discuss now any conclusions which can be made for the
case of real galactic nuclei and environments from our results. This will be
useful for observational programmes on TDR.
To predict the TDR in real galactic nuclei, we use the method of scaling. The TDR
obtained in our simulations has to be scaled up in two ways: first from relatively low ($N\sim 10^{5}$)
to more realistic high particle numbers ($N \sim 10^{8}$). Second,
our accretion radius $r_t$ has been chosen very large compared to any realistic tidal radius
(for smaller $N$ simulations it has to be done in order to get any meaningful results on TDR).
So, we also have to discuss how to scale down the TDR from our simulated values of $r_t$
($10^{-3}, 10^{-4}$) to the small more realistic regime of $r_t$ ($10^{-7}$).
This can be done by applying scaling relations from known scaling laws (obtained e.g. from
Paper I and other literature) for $N$, and an empirically determined one for $r_t$. We have
shown above that the TDR in King ($W_0=6$) and Plummer models
is very similar, so we expect the scaling formula (A10) derived in Paper I for a Plummer model
to be also valid for our King model used here.
So, we apply the same boost factor of TDR with respect to the axisymmetry of a galactic nucleus
for the real galactic nucleus as we find here in this paper for our simulated systems.
For example, in Paper I we estimated the TDR
of the Milky Way SMBH to be $1.09\times10^{-5} \rm yr^{-1}$ after a scaling procedure with respect
to $N$ and $r_t$. By fitting surface brightness
profile to mid-infrared images of the nuclear cluster in our Milky Way, \citet{SFK2014}
reported the mean ratio between minor and major axes is 0.71, which is close to
our rotating $\omega_0=0.6$ model. For this model we find a boost of TDR by 35\% in our simulations,
and we apply the same factor here for the case of axisymmetry, to get a higher TDR of
$1.47\times10^{-5} \rm yr^{-1}$.

\subsection{Relation to other current papers in the field}

With regard to the enhancement of TDR in axisymmetric systems we have shown that our
results are in agreement with \citet{FPB2012}; but recently numerical simulations
published by \citet{VM2013} and \citet{Vasiliev2014} seem to contradict our findings.
They claimed that the TDR in axisymmeric nuclei can be a few times larger than in the spherical case.
Also \citet{LHBK2014} analyzed the distribution of stellar orbits in an axisymmetric galaxy and
found that total number of stars that can interact with the central SMBH binary is
six times larger than in the spherical system. In this subsection we will discuss why
there is such a discrepancy to our results - we find a much smaller enhancement of
TDR in axisymmetric systems.

The main difference between the cited papers and our work is the initial model. In all of the
above mentioned papers, a flattened Dehnen model is used (their density profile,
given the parameters they chose, is identical also to the Hernquist model).
Their models possess a fixed axial ratio ($c/a = 0.75$) throughout
the entire cluster and an initial central cusp, while our rotating initial model has initially a core
density distribution in the center, and we have a
radial variation of $c/a$ from nearly spherical ($c/a\approx 1$) to about $c/a \approx 0.7$ in
the outskirts (see Fig.~\ref{fig_AxisRatio}).
However, in the radius range where most of the disrupted stars originate from
(c.f. Fig.~\ref{fig_Rmax2D_R}), the system deviates significantly from spherical symmetry, thus
we can confirm that the enhancement of TDR is connected with the non-spherical geometry.
But in the relevant region of our $\omega_0=0.6$ deviation from spherical symmetry is
less than in the other cited papers, which may be an explanation for the weaker effect in
our case.

We also notice that even within the cited other papers there are some discrepancies in the results
even for models with the same initial density profile.
For example, the enhancement of the number of accreted stars in \citet{VM2013} was
smaller than 100\% (see Table 2 in their paper), while \citet{LHBK2014} found a factor of six.
On the other hand some of the models in \citet{VM2013} only show mild
enhancement which is in the same level as ours. Another example comes from the debate about the
``final parsec problem'' in SMBH binary evolution. Based on their simulation results, \citet{KHB2013}
claimed that the ``final parsec problem" is not a problem in axisymmetric host galaxies,
while \citet{VAM2014} reached an opposite conclusion according to their simulation. We
notice that both of these work employs similar flattened galaxies model, however, they
used a different method to generate the initial model.

\citet{VM2013}, \citet{Vasiliev2014} and \citet{VAM2014} utilized the orbital superposition
method of \citep{Schwarzschild1979} to construct their model. On the other hand, \citet{KHB2013} and \citet{LHBK2014}
used another method called ``adiabatic squeeze technique" developed by \citet{HBMS2001}.
We notice that in the process of adiabatic squeeze, which contains a step which applies a slow and smooth velocity
change on the stars in the $z$ direction. This step may artificially reduce the energy and angular
momentum of the stars in the model cluster. Although the radius and velocity vectors of the stars
are rescaled after the squeeze, it is not clear how the rescaling affects the phase space distribution.
Thus it might be possible that the process produces more stars of low energy and low angular
momentum. Another evidence of a similar effect can be derived from \citet{Vasiliev2014}; while they
still use the orbital superposition method they changed the generation of their initial model so
that it creates more low energy and low angular momentum stars. In their test run (Fig. 2 in their paper)
we see a much larger enhancement of the number of accreted stars compared to \citet{VM2013}.
So to add more low energy and angular momentum stars seems to be promising in abridging the different
enhancement factors between \citet{VM2013} and \citet{LHBK2014}. We suggest that a detailed
comparison between models constructed with these two methods (and their phase space distribution) should
be performed in order to explain the discrepancy.

According to \citet{LHBK2014} the central two parsecs of their model galaxy exhibit a slight triaxiality, which
could also introduce some additional centrophilic orbits, thus increase the number of stars that can interact with
the central SMBH binary.

Before finishing this section, we want to make a final remark on the result of \citet{LHBK2014}. Their model
integrates individual orbits in a fixed model potential with one SMBH
in the center, in a static way. So the number of stars that can interact with the central SMBH binary according
to their results should be
considered as an upper limit. Once two-body relaxation is turned on, some of the stars that are
supposed to be inside the loss cone might be scattered out. And the presence of a SMBH binary in an evolving
system may also affect the result of how many stars can interact with them.

%%%%%%%%%%%%%%%%%%%%%%%%%%%%%%%%%%%%%%%%%%%%%%%%%%%%%%%%%%%%%%%%%%%%%%%%%%%%%%%%%%%%%%%%%%%%
\section{Loss cone in axisymmetric potential}

First, we summarize the loss cone theory for stellar orbits in a spherically symmetric gravitational
potential, in order to discuss different behavior in an axisymmetric potential later.
If a stellar orbit has a pericenter distance less than the tidal radius it is considered to be in the loss cone.
In  spherical symmetry the boundary of the loss cone can be expressed in terms of a critical loss
cone angular momentum
$J_{lc} \approx \sqrt{2 G M_{\bullet} r_t}$ (if $r \gg r_t$; cf. e.g. \citet{AFS2004}).
The loss cone is then defined as the region in phase space where the angular momentum $J$ of a star
fulfils $J < J_{lc}$. All stars inside the loss cone will reach the tidal radius within
a dynamical (orbit) time scale. As a consequence the loss cone would become empty in that
relatively short time. Once a star is inside the loss cone and reaches the tidal radius,
we assume that it will be destroyed by the BH's tidal force instantaneously and add its total mass
to the black hole at the same moment. Most authors studying stellar dynamics and TDR of star clusters around a BH
used similar approximations. Rees (1988) already argued that the stellar debris after
tidal disruption will make several orbits until it is finally accreted by the BH; nevertheless
the orbital time near the BH is very short compared to the original orbital time of the star before
its disruption.
Recent detailed simulations on tidal disruptions \citep{GRR2013,HSL2013,HSL2015}
show that in some case not all material of the star may be accreted and that
general assumptions about the tidal fallback rate are not correct; for example
in a longer lived accretion disk may form, which would delay the black hole growth.
In a spherical system, without interactions between the stars, angular momentum $\bf{J}$ would
be strictly conserved. So,
without any repopulation of the loss cone, the accretion process would stop after a few dynamical
times. But stars do interact with each other while moving inside the star cluster by two body relaxation
through mutual encounters; in this process they can exchange angular momentum and energy and
so the loss cone will be repopulated in the two body relaxation time scale, which is generally
long compared to the dynamical time \citep{CK1978,AFS2004}. The repopulation of the loss cone is modelled
in these papers as a diffusive process using the Fokker-Planck approximation.

In an axisymmetric potential, the situation is more complex since $\bf{J}$ is not a conserved quantity. It changes
continuously due to the non-central force resulting from the geometry of the potential. In this case, stars
with $J > J_{lc}$ may have a chance to drift into the loss-cone and get disrupted. In other words, the loss cone
is enlarged in the $J$ dimension in axisymmetric potential. However, the $z$ component of angular momentum $J_z$
is still conserved, so a solid boundary of the loss-cone is $J_z \leq J_{lc}$. \citet{MT1999} investigated
this topic using a symplectic map introduced by \citet{TT1997}. In this work, we analyze the enlarged
loss cone in phase space in terms of energy $E$, modulus of angular momentum $J$ and the $z$ component of
angular momentum $J_z$ for stellar orbits near the BH. We use a different approach as \citet{MT1999}
here, which is based on a numerical particle scattering experiment.
In what follows, we first describe the method we used in this experiment, then present our results.

First step, we need to know the smooth gravitational potential as a function of position without
the fluctuations due to the discrete particle structure. We use a so-called self-consistent
field code (SCF, \citet{OH1992}) to generate the analytical function for the gravitational potential.
The expansion coefficients
$C_{lm}, D_{lm}, E_{lm}, F_{lm}$ used in computing forces (Eq.(3.21)-(3.23) in \citet{OH1992})
are computed based on snapshot data generated during the direct $N$-body simulation. By default, the code
uses radial basis functions labeled from $n=0$ to $n_{max}=14$, and spherical harmonic function truncated at $l_{max}=10$.

Our particle distribution is self consistently achieved as a consequence of the
co-evolution of stars and BH. Using the SCF code means that all two-body interactions are smoothed out in the
experiment. Because we assume that most of the two-body interactions happens during the apocenter passage,
which is also used in \citet{TT1997}. After getting the coefficients, we can calculate the acceleration,
jerk and do orbit integration using a Hermite $4^{th}$ integrator with variable time steps, developed by
ourselves. This code works very well and the energy and angular momentum errors of the test particle stays
in the level of $10^{-9}$ over long time integration. In an axisymmetric system all coefficient with $m\neq 0$
should be 0. But in practice one will get some small numbers very close to 0 due to particle noise. We just
ignore these terms, otherwise $J_z$ would no longer be conserved. We also ignore coefficients with
odd $l$, because the rotating system should be symmetric about the equatorial plane and do not have
pear-like shape.

Next step is to generate initial positions and velocities for test particles. The basic idea of this experiment
is to do parameter space scanning. We uniformly sample $E, J$ and $J_z$, all test particles are initially put
at their apocenter. Firstly, we choose a particular energy and calculate $J_{lc}$ through equation
$J_{lc} = r_{t}\sqrt{2(\Phi(r_t)-E)}$. Then we choose a pair of $(J,J_z)$, $J$ can be a few times larger than
$J_{lc}$ but $J_z$ keeps smaller than $J_{lc}$. Given the combination of $(E,J,J_z)$ and the potential
distribution we can find the apocenter position given by $(r,\theta)$. Here $r$ is distance to center
and $\theta$ is the angle between position vector and $z$-axis. We note that there
are actually four parameters $(E,J,J_z,\theta)$ to define the initial conditions for a particular orbit.
So we further sampled 100 data points in $\theta$ dimension. In order to plot the result in a
2D plane, we introduce a filling factor $P$ for every $(E,J,J_z)$ combination to describe this $\theta$
dependence, which is the fraction of stars in the loss cone for a given combination of
$(E,J,J_z)$ (number of data points in loss cone divided by total sample size, e.g. 100), meaning that among
all stars with same $(E,J,J_z)$ only a fraction of $P$ are inside the loss cone.

By our definition a star in the loss cone will be disrupted by the BH within one dynamical time,
so for every test particle
we only integrate their orbits for one orbital cycle. If a particle comes back to its apocenter,
we consider it as out of the loss cone and move to the next integration with new initial orbital data.

\begin{figure}[htbp]
  \begin{center}
  \includegraphics[width=\columnwidth]{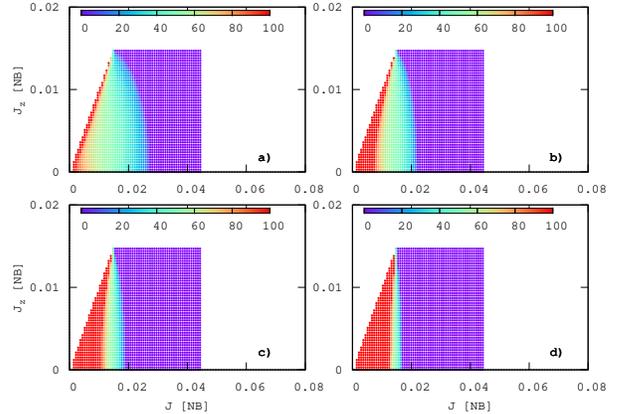}
  \end{center}
  \caption{Filling factor $P$ of the loss coneas a function of $J$ and $J_z$ for different energies;
the $x$ axis is the modulus of the total angular momentum $J$, while the $y$ axis is its $z$ component $J_z$.
Panel a) correspond to $E=-1.3$, b) $E=-1.5$, c) $E=-1.7$, d) $E=-1.9$. Colors represent the
filling factor $P$ in percentage. All data are given in standard $N$-body units.}
  \label{fig_EJJz-multi}
\end{figure}

Fig. \ref{fig_EJJz-multi} shows results from the experiment in a slowly rotating model ($\omega_0 = 0.3$),
it represents the loss cone shape in phase space. Since $J$ is not conserved we use its
initial value at apocenter for the figure;
at the time of disruption $J$ must be less than $J_{lc}$. From panel a) to d), the energy of the test
particles are in descending sequence, so their position of apocenters are getting closer and closer to
central BH. One can see that the whole plane comprises 3 regions: 1) inner region where $P$ equals 1,
meaning particles with these $(E,J,J_z)$ can hit the BH within one dynamical time scale; 2) transition
region where $P$ is non-zero but less than 1, particles with these $(E,J,J_z)$ have a chance to hit the BH
depending on their apocenter position ($\theta$ value); 3) outer region where $P=0$, none of particle in
this region can hit the BH. In panel a) one can see only a few points are red and a lot of points are
located in transition region. From a) to d), the fraction of $P=1$ points in the $(J,J_z)$-plane
increases and the transition region is compressed
by the inner and outer region in horizontal direction ($J$ dimension). This is because test particles with
high energy (loosely bound or unbound with respect to the BH) can go beyond the BH's influence radius
to the intermediate and outer regions of the cluster, where the axisymmetric
stellar potential dominates. The angular momentum of these test particles will have large variations.
So a wide transition region exists in high energy cases. But in the low energy case (stars
strongly bound to the BH), e.g. panel d),
test particles are moving inside the BH's influence sphere where the potential is dominated by the
BH and thus approaches spherical symmetry. All loss cone stars following the classical loss cone
approximation, should have both $J$ and $J_z$ to be smaller than $J_{lc}$. In all panels of Fig. \ref{fig_EJJz-multi},
on the contrary, we see how stars with $J > J_{lc}$ could be still in the new, extended loss cone of an
axisymmetric system with a certain non-zero probability.

\begin{figure}[htbp]
  \begin{center}
  \includegraphics[width=0.47\columnwidth]{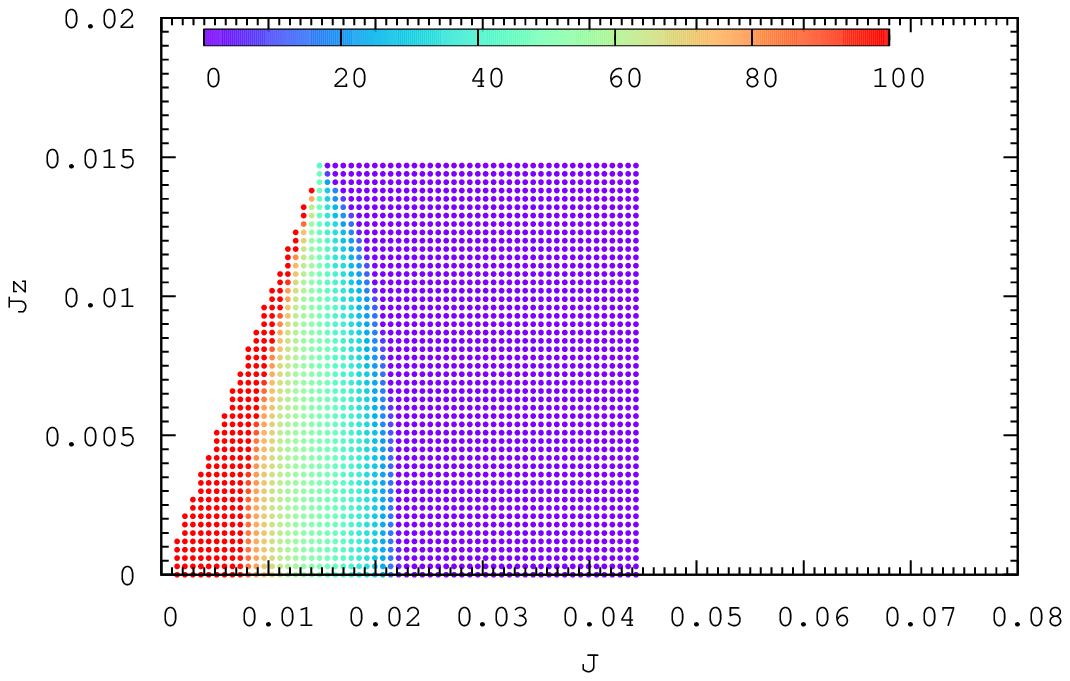}
  \includegraphics[width=0.47\columnwidth]{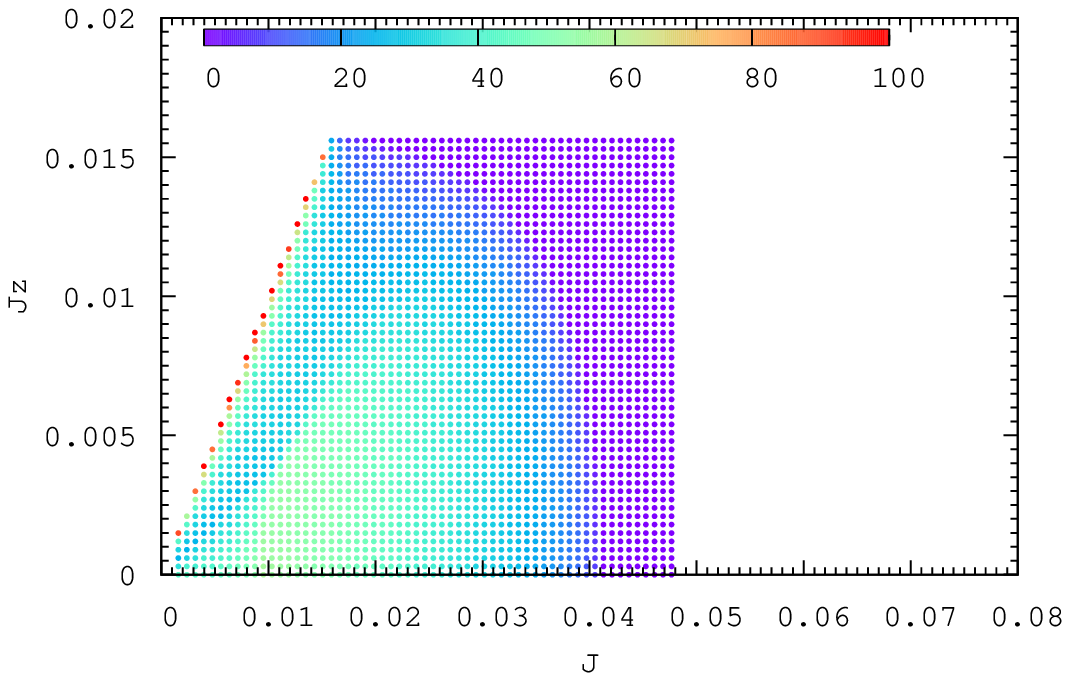}
  \end{center}
  \caption{Comparison between 2 models with same test particle energy $-1.5$. $x$ axis is module of
angular momentum in $N$-body unit. $y$ axis is $z$ component of angular momentum. Left panel correspond
to $\omega = 0.3$; right panel $\omega = 0.6$. Colors indicate the filling factor $P$ in percentage. }
  \label{fig_EJJz-compare}
\end{figure}

For faster rotating models ($\omega_0 = 0.6$) the results are similar. Three regions are presented on
the $(J, J_z)$ plane, however, the extent of each region is different from the counterpart of same energy
in slow rotating model. Fig. \ref{fig_EJJz-compare} gives an example, in both left ($\omega_0 = 0.3$) and
right ($\omega_0 = 0.6$) panel the test particle have same energy, however, the resulting appearances are
quite different. In the left panel we see the the outer border extended to $J = 0.024$, while in the right one the
outer border goes to $J = 0.04$ and is not as clear as that in the left panel. Also in the right panel the
red region is almost disappeared. These results show how rotation modifies the loss cone shape
in phase space. In both of these plots, the maximum radius stars can achieve are roughly the same.
However, faster rotation means we have a more flattened cluster shape, which enhances the torque acting on stars,
thus the variation in $J$ becomes larger. So, the higher the degree of rotation in the stellar system,
the larger is the extension of the loss cone in $J$ direction.

On first glance at Fig.~\ref{fig_EJJz-multi} (also Fig.~\ref{fig_EJJz-compare}) one might think that
the loss cone is generally enlarged by a significant factor.
However, as we pointed out above, there is a filling factor $P$ for every point
on the $(J,J_z)$ plane. To find the net enlargement of the loss cone in axisymmetric potentials
we introduce an effective area $S$ of the loss cone in these plots
by integrating the filling factor $P$ over the $(J,J_z)$ plane.
For example, the effective area of the classical loss cone is just given by the size of the triangle $J,J_z < J_{lc}$
in our plots, since in the classical case $P$ is unity everywhere in this triangle region.

\begin{figure}[htbp]
  \begin{center}
  \includegraphics[width=\columnwidth]{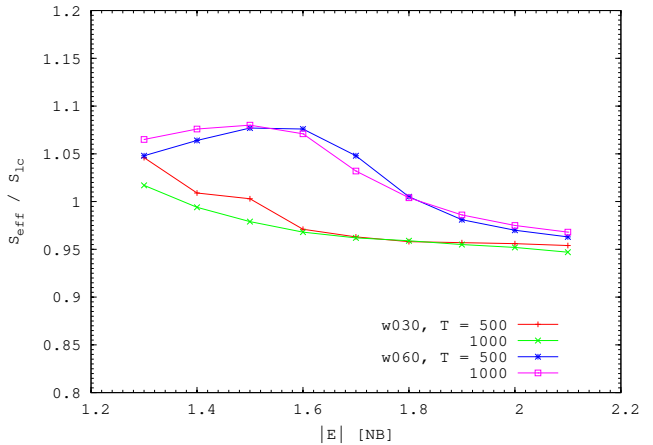}
  \end{center}
  \caption{Ratio between effective area $S_{eff}$ of loss-cone in axisymmetric system and $S_{lc}$ in
spherical system.}
  \label{fig_EffArea}
\end{figure}

Now we compare the loss-cone size comparing the integrals $S$ with each other. We define the quotient
$\alpha_{lc} = S_{eff}/S_{lc}$, where $S_{lc} = J_{lc}^2/2$ is the classical loss cone integral.
$\alpha_{lc}$ is plotted in Fig.~\ref{fig_EffArea} as a function of binding energy $|E|$.
In the plot we show both slow and fast rotating models at two different evolution times. For the slow rotating model
the ratio $\alpha_{lc}$ is even smaller than unity for binding energies larger than 1.4 - 1.5, meaning that
at large $|E|$ the loss cone is smaller compared to the classical one. This is caused by the reduction
of the probability $P$ at the boundary and inside the classical loss cone region $J \le J_{lc}$.
$P$ is decreasing from inside toward outside. While for intermediate $|E|$ case,
although $P$ is still decreasing function of $J$, the large number of valid points overwhelms, so the net
effect is increasing the effective area. However, if one goes further toward small $|E|$ the ratio will
drop again, like the case of fast rotating model. This is just because $P$ is sufficiently small in this
case and win the game. For fast rotating model, another interesting feature is the ratio drops below 1 at
$|E| = 1.8$.
From this figure we see that the enlargement of loss cone, quantified by the ratio
$\alpha_{lc}$ as a function of binding energy.
Interestingly, the change of the effective loss cone size in every energy slice is less than 5-10\%.
These mild changes seem unable to raise TDR with the amount observed in the simulation, to address this it will
be useful if we can estimate the TDR based on the effective loss cone measurement and compare with simulation.
However, the knowledge of how stars are distributed in energy and angular momentum is required. With the limited
particle numbers of the model cluster, it is difficult to get an accurate and reliable distribution function. Also in
current work, we sample the energy space with large intervals ($\Delta E = 0.1$), which may cause large errors in the
estimated TDR. So we did not make the estimation. There are still plenty of works could be done with this topic.

%%%%%%%%%%%%%%%%%%%%%%%%%%%%%%%%%%%%%%%%%%%%%%%%%%%%%%%%%%%%%%%%%%%%%%%%%%%%%%%%%%%%%%%%%%%%

\section{Orbital properties of disrupted stars}

In this section, we investigate the origin of disrupted stars. Under the assumption that stars in
loss cone can survive for only one orbital period, the origin of these stars can be
examined by looking at their energy and angular momentum, as well as their origin (apocenter)
in spatial coordinates (radius and angle $\theta$).
In spherical systems one can use effective potential to compute the apocenter of orbit, but
in the axisymmetric case we do not have such convenient solutions except to run the simulation twice.
In the first run we find out the ID for those stars that will be disrupted by BH. Then, in the second run
we make records for these stars more frequently than other stars, in order to catch their last apocenter position.

\begin{figure}[htbp]
  \begin{center}
  \includegraphics[width=\columnwidth]{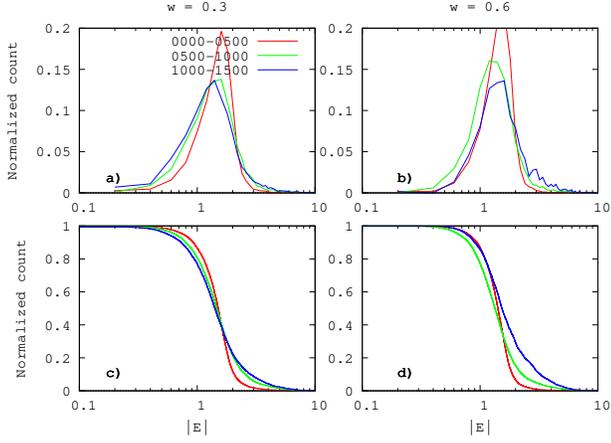}
  \end{center}
  \caption{Panel a) and b) show normalized distribution of binding energy $|E|$ of tidally disrupted stars, for different rotating models and for three different time intervals (indicated by color) in the full $N$-body simulation with $r_t = 10^{-3}$. The distribution is normalized to the total number of disrupted stars in each time interval. Panel c) and d) show cumulative fraction profile corresponding to a) and b), respectively.}
  \label{fig_E_stat}
\end{figure}

\begin{figure}[htbp]
  \begin{center}
  \includegraphics[width=\columnwidth]{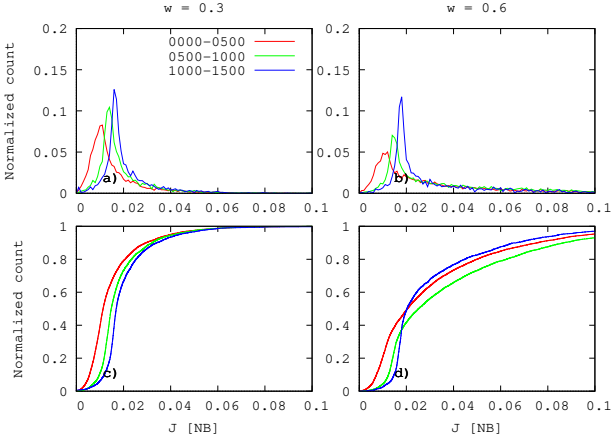}
  \end{center}
  \caption{Same as Fig.~\ref{fig_E_stat}, but here for the distribution of total angular momentum of the disrupted stars.}
  \label{fig_J_stat}
\end{figure}

We found in the beginning that the total TDR, especially for small $r_t$ does only marginally
depend on the rotation of the system; consistent with this we found in the previous chapter
that the loss cone structure does change significantly, but the total integral over the loss
cone space for axisymmetric systems yields only relatively small changes.
Still it is interesting to study how the orbital properties of stars, which are tidally disrupted,
change with the rotation of the system.
In order to address this, we now turn back to our full $N$ body simulations and study
the distribution of $|E|$ (Fig. \ref{fig_E_stat}) and $J$ (Fig. \ref{fig_J_stat}) of the disrupted stars at their apocenter passage in three time intervals. From Fig. \ref{fig_E_stat} one can see that most of the tidally disrupted stars have a binding energy between 1 and 2, coincident with the small bumps in Fig.~\ref{fig_EffArea} where $\alpha_{lc} > 1$. Another
evidence comes from the distribution of $J$ as shown in Fig.~\ref{fig_J_stat}, where one can see the peaks
are lying outside of the $J_{lc}$ which is roughly 0.015. The peaks are moving toward larger $J$, which is
caused by the increase of BH mass (recall the expression for $J_{lc}$). A significant fraction of stars
comes from places outside of the classical loss cone in $(J,J_z)$ plane.

\begin{figure}[htbp]
  \begin{center}
  \includegraphics[width=0.9\columnwidth]{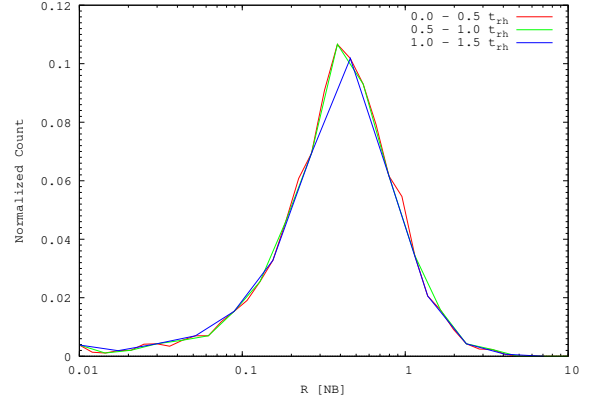}
  \end{center}
  \caption{Distribution of last apocenter distance of disrupted stars in 3 different time interval. Each curve are normalized to the total number of disrupted stars in given time interval.}
  \label{fig_Rmax2D_R}
\end{figure}

\begin{figure}[htbp]
  \begin{center}
  \includegraphics[width=0.47\columnwidth]{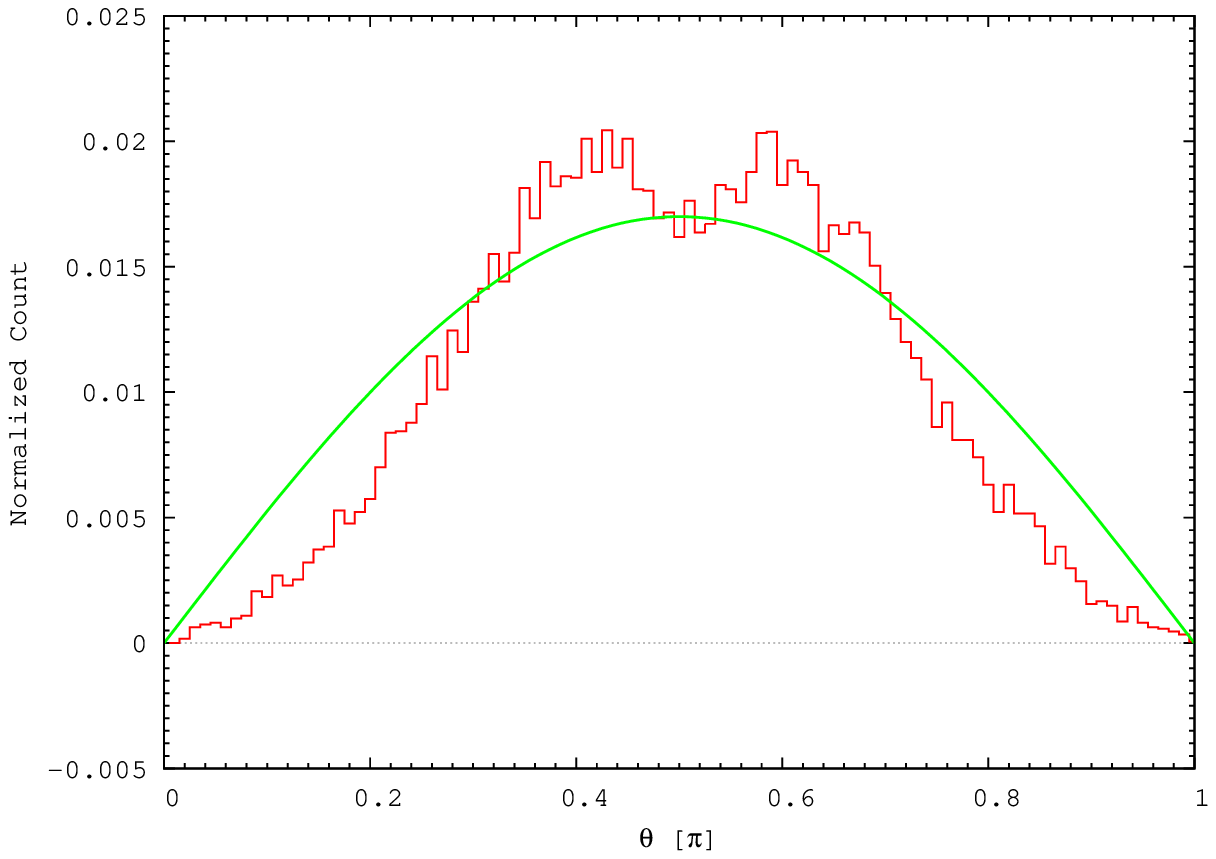}
  \includegraphics[width=0.47\columnwidth]{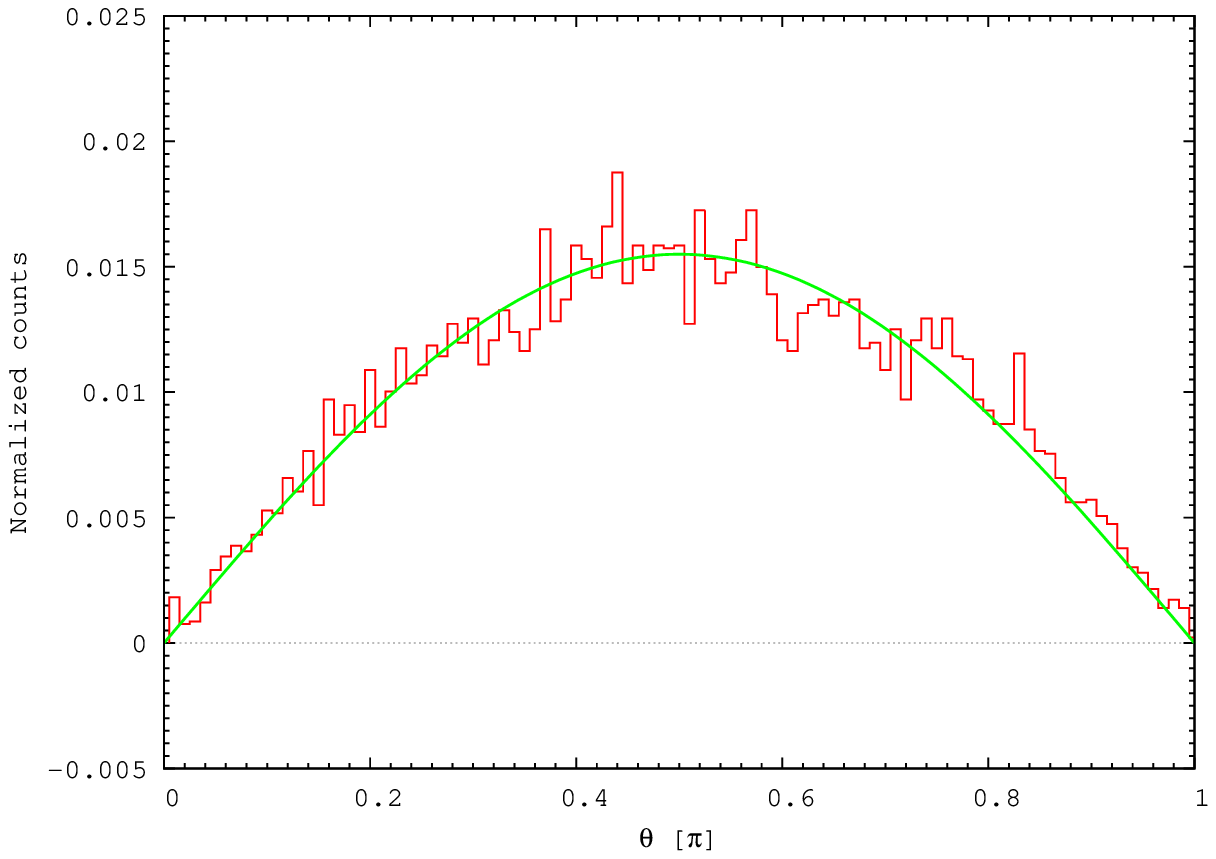}
  \end{center}
  \caption{Normalized distribution of zenith angle $\theta$ of last apocenter of disrupted stars. Left panel is axisymmetric model, right panel is spherical model.}
  \label{fig_Rmax2D_T}
\end{figure}

\begin{figure}[htbp]
  \begin{center}
  \includegraphics[width=0.9\columnwidth]{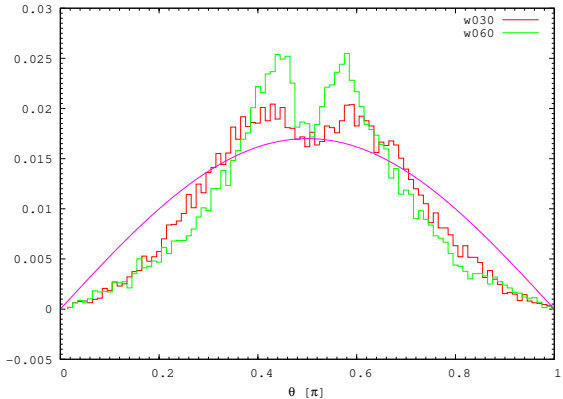}
  \end{center}
  \caption{Normalized distribution of zenith angle $\theta$ of last apocenter of disrupted stars.
  Compare between slow (w030) and fast (w060) rotating models.}
  \label{fig_Rmax2D_T_comp}
\end{figure}

In spherical systems it is usually sufficient to describe the apocenter of an orbit by its radial
distance from the center (the BH); the orientation of the orbit does not play any role for the orbital
time and the nature of the encounter with the central BH. However, in axisymmetric systems, orbits with
different angle $\theta$ (the angle between position vector of the star at apocenter and the $z$-axis)
will differ from each other significantly. Therefore we have to describe the distribution of apocenters
of tidally disrupted stars in terms of both the
$r$ (Fig.~\ref{fig_Rmax2D_R}) and $\theta$ (Fig.~\ref{fig_Rmax2D_T}) dimension. From
Fig.~\ref{fig_Rmax2D_R} one can see the peaks are quite far from the BH, in a region
comparable to the BH influence radius, which is similar to the apocenter
distribution in spherical systems (Paper I). The difference turns out to be in the $\theta$ dimension,
as shown in Fig.~\ref{fig_Rmax2D_T}. We compare the $\theta$ distribution between spherical and
axisymmetric systems. Imagine we project all the apocenter points onto a sphere with radii equals 1.
The measured number counts in each $\theta$ bin $\Delta N(\theta)$ are computed by
$2\pi\cdot\Sigma(\theta)\cdot\sin(\theta)\Delta\theta$, where $\Sigma(\theta)$ is the surface density
of projected points on the unit sphere. If apocenters are uniformly distributed with $\theta$,
$\Sigma(\theta)$ is constant, then $\Delta N(\theta) \propto \sin(\theta) \Delta\theta$. Here we choose
an equal bin size, so the measured number count should follow a $\sin(\theta)$ curve. The right panel of
Fig.~\ref{fig_Rmax2D_T} plots $\theta$ distribution for spherical model, which is taken from our last
work (Paper I). In left panel we see the last apocenter distribution have deficit at polar region comparing
to $\sin(\theta)$ curve, and excess at places beyond and below the equatorial plane, showing a double peak
feature. The deficit at the polar region may have something to do with the flattening of the cluster, however, this
is not the only reason. The double peak feature around the equatorial plane obviously does not relate to a
geometrical origin, otherwise the peak should be placed at the equatorial plane. In Fig.~\ref{fig_Rmax2D_T_comp}
we compare the $\theta$ distribution between slow and fast rotating models. One can see that in fast
rotating model, the double peaks are more significant, accompanied by a further deficit in the angle range
$0.2 - 0.4 \pi$ and $0.6 - 0.8 \pi$.

In order to understand the double peak feature, we turn to the orbit structure of these disrupted stars.
In non-spherical symmetric stellar system with a SMBH in its center, the space populated by stars can be
divided into three parts depending on the distance to the BH, namely the regular, chaotic and mixing region
\citep{PM2001}. Inside the BH's influence radius $r_h$, the potential felt by the star is dominated by the BH plus
a small perturbation from the non-spherical stellar potential. In this region, the motion of stars is
essentially regular, as in a spherical potential. Outside of $r_h$, stars passing the center
will suffer a large angle deflection by the BH, which in conjunction with the non-spherical potential
near and outside $r_h$, could make their orbits stochastic.

We are interested in stellar orbits in an axisymmetric stellar potential, which can get close to the central BH. These are typically two classes of orbits, short-axis tube (SAT) and saucer (see \citet{Vasiliev2014} for example); they can be distinguished by their third integral of motion $I_{3}$. Although $I_{3}$ may help us quickly distinguish orbit families, finding the functional form of $I_{3}$ is difficult (see \citet{LG1987} and discussion in \citet{ST1999}) and is beyond the scope of this paper. We choose alternative ways to do orbit classification, such as Surface of Section (SoS) plot and Fourier analysis of $J_x$ (see Appendix).

\begin{figure}[htbp]
  \begin{center}
  \includegraphics[width=0.47\columnwidth]{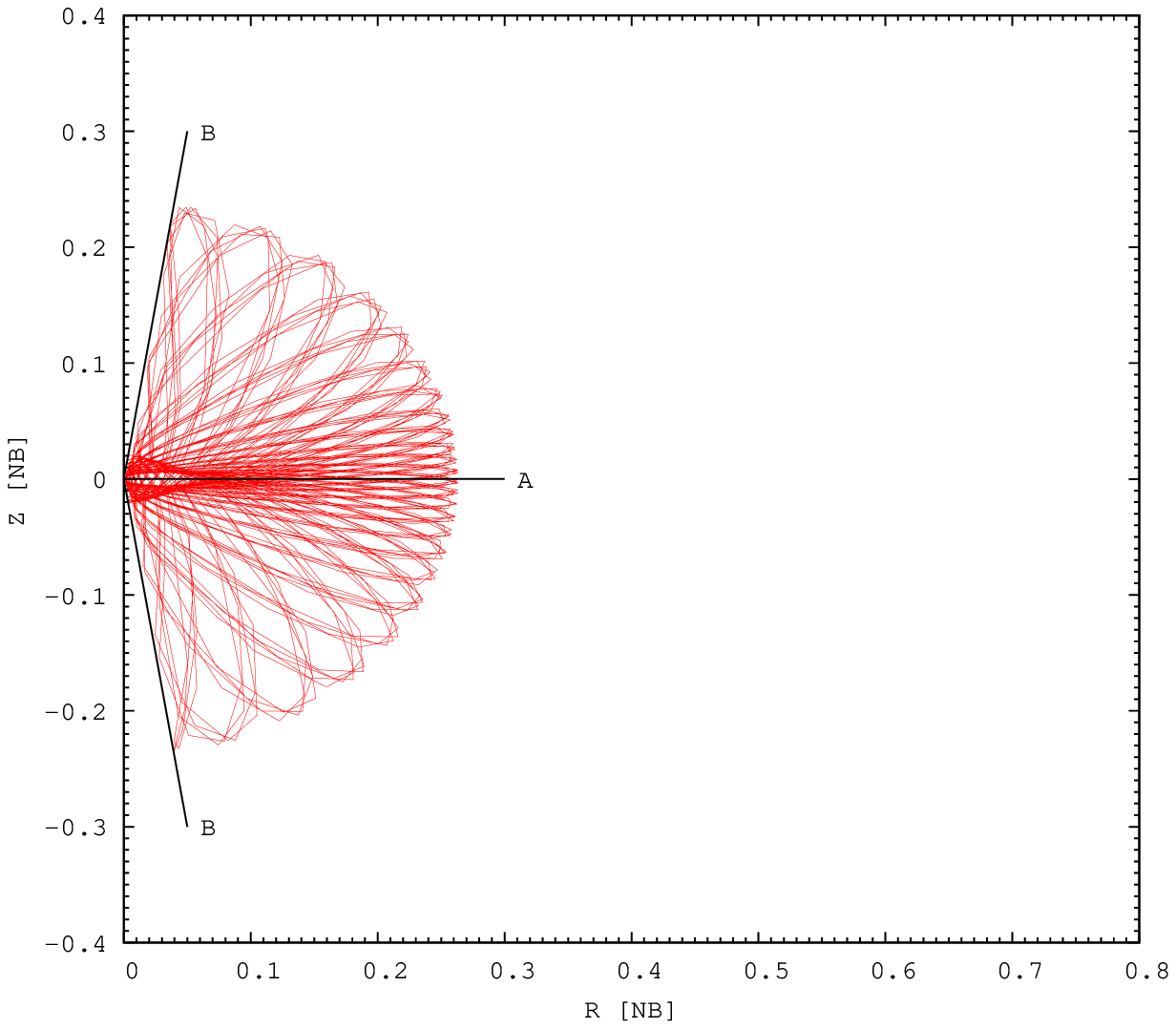}
  \includegraphics[width=0.47\columnwidth]{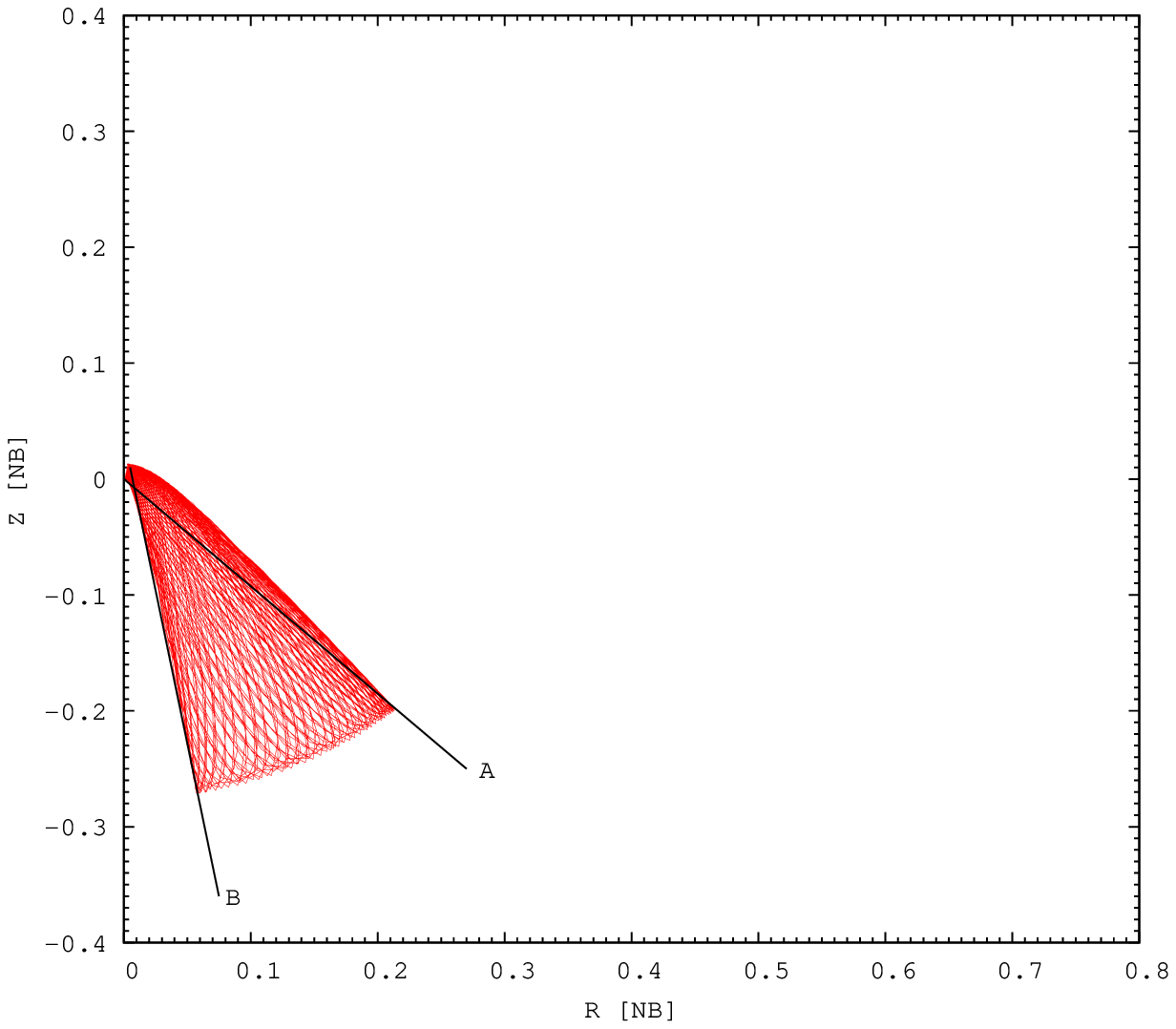}
  \end{center}
  \caption{Examples of orbital structure for SAT (left) and saucer (right) orbit. Stars achieve maximum
  $J$ when their instant orbital plane coincide with \textbf{B} plane, while minimum when coincide with
  \textbf{A} plane.}
  \label{fig_Orbit_R-z}
\end{figure}

\begin{figure}[htbp]
  \begin{center}
  \includegraphics[width=0.9\columnwidth]{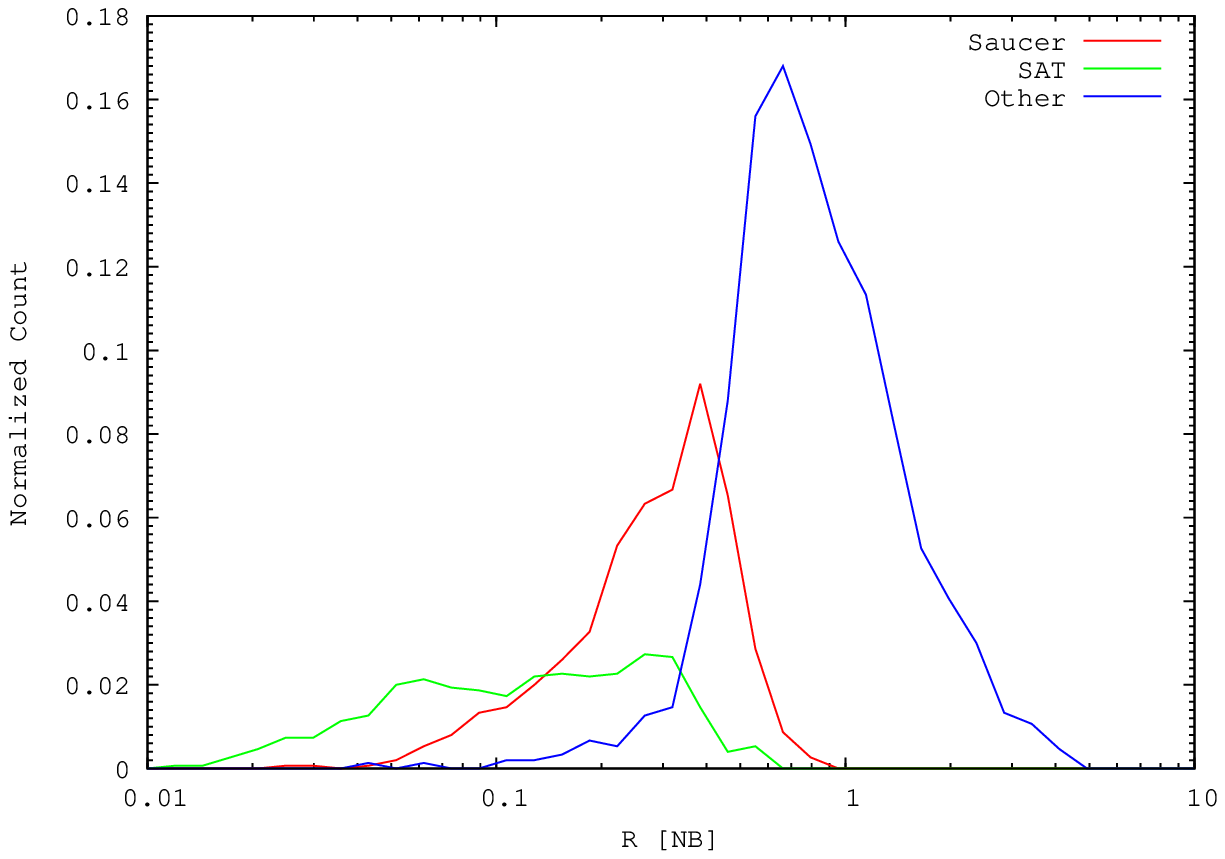}
  \includegraphics[width=0.9\columnwidth]{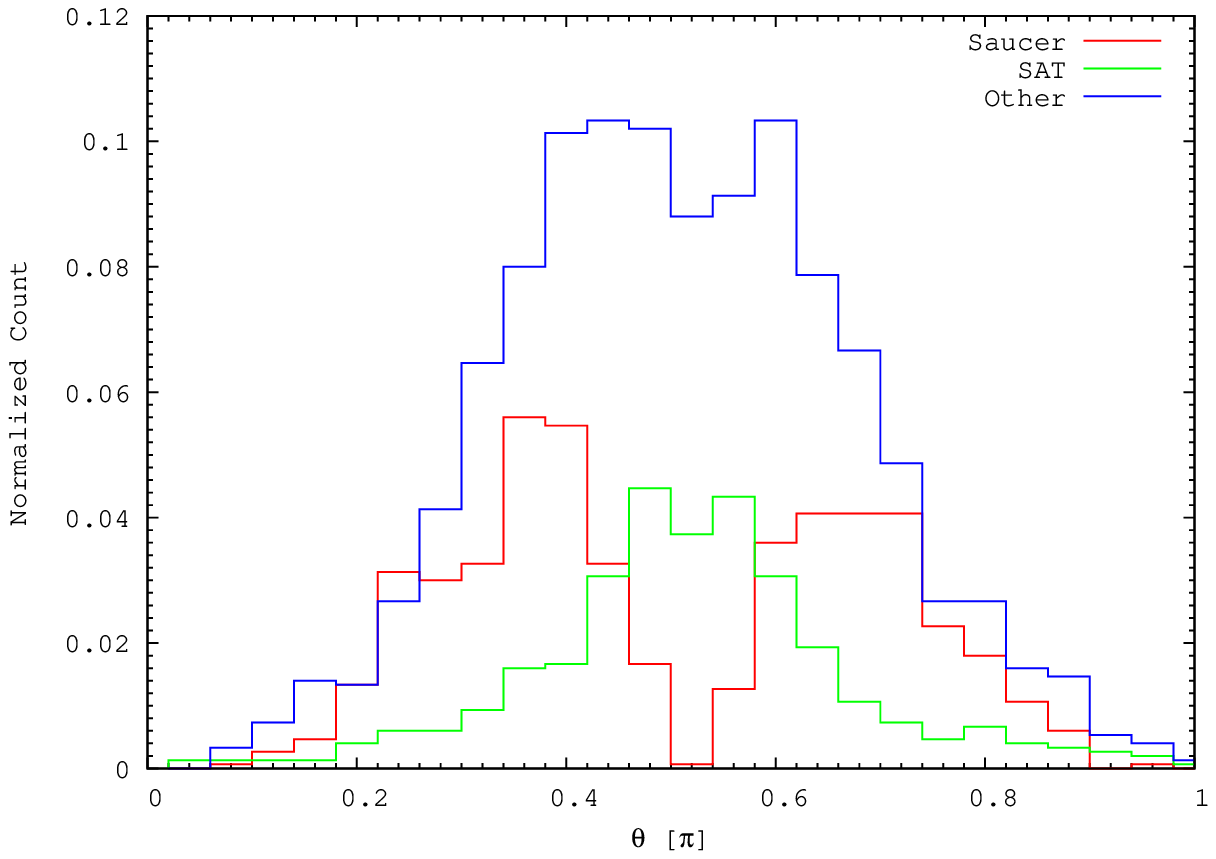}
  \end{center}
  \caption{Normalized distribution of apocenter distance $r$ and zenith angle $\theta$ of last apocenter of disrupted stars for different orbit families. The distribution is normalized to the number of stars in each orbit family.}
  \label{fig_Rmax2D_Class}
\end{figure}

Fig.~\ref{fig_Orbit_R-z} gives examples of SAT and saucer orbits in configuration space. The plot is made
in cylindrical coordinates so that one can catch the main point easily. For SAT orbit, one can see its
apocenter can go both above and below the equatorial plane. While apocenter of saucer orbit can only exist
on one side of mid-plane, due to restrictions by the 3rd integral.
We also check the value of $J$ at each apocenter passage. We find that SAT orbit
achieve its minimum $J$ at the equatorial plane; a saucer orbit cannot reach the
equatorial plane, but its minimum $J$ is achieved at the place which is next to the equatorial plane as marked
in the plot by \textbf{A} plane. Recall in the last section we said no matter what $J$ one star has at the
apocenter, at the time of disruption it must be smaller than $J_{lc}$. So the last apocenter place should
be around the \textbf{A} plane. This seems to be promising to explain the double peak in $\theta$ distribution,
however, need to be confirmed. In order to see this we try to do orbit classification for the disrupted
stars, which is computationally expensive. So we just randomly select a sub-sample of disrupted stars
and divide them into 3 orbit families: SAT, saucer and others (here ``others" means they do not belong to the
former two families, and may be chaotic orbits). Among the 2943 sample stars, 1719 are classified as
``others'', 757 as saucer and 467 as SAT. Then we re-plot the $r$ and $\theta$ distribution for different
orbit families in Fig.~\ref{fig_Rmax2D_Class}.

The results show that the apocenter distribution of different orbit families not only differs in $\theta$ but
also in $r$. One can see the innermost region is dominated by SAT orbits, and concentrated to the equatorial
plane. Intermediate radius is mostly occupied by saucer orbits, and the distribution in $\theta$ shows
double peaks as expected. Further out is the region dominated by orbits marked as others. These orbits
can go outside of the influence radius and are basically chaotic orbits. From Fig.~\ref{fig_Rmax2D_Class}
one can also find out the fractions of each orbit family contributing to the budget of disrupted stars:
the largest fraction comes from chaotic orbits; SAT orbits contribute least to the budget because they are
deeply buried in the cluster center where the total star number is small; the intermediate contribution is from
saucer orbits which create the two peaks in the $\theta$ distribution.

%%%%%%%%%%%%%%%%%%%%%%%%%%%%%%%%%%%%%%%%%%%%%%%%%%%%%%%%%%%%%%%%%%%%%%%%%%%%%%%%%%%%%%%%%%%%
\section{Conclusions and Discussions}

Tidal Disruption (TD) of stars by supermassive central black holes (SMBH) from dense rotating star
clusters is modelled
by high-accuracy direct $N$-body simulation. As in a previous paper on spherical star clusters we study the
time evolution of the stellar tidal disruption rate and the origin of tidally disrupted stars, now according to
several classes of orbits which only occur in rotating axisymmetric systems (short axis tube and saucer).
In empty loss cone regime, comparing spherically symmetric and axisymmetric systems we find a higher TD rate in large $r_t$ models in axisymmetric case, but for small $r_t$ case - somewhat surprisingly - there is virtually no difference in the TD rate, maybe a small increase due to axisymmetry.

We define an extended loss cone by the condition that stars in the axisymmetric potential reach the BH
within one orbit. A detailed analysis shows that the structure of the loss cone significantly differs
from the spherical case; if $J_{lc}$ is the critical angular momentum to be in the loss cone in a
spherical system, and $J,J_z$ are the total and $z$-component of the angular momentum of a stellar
orbit, there are many stars with $J > J_{lc}$ in the loss cone; since, however, there are also
some stars with $J > J_{lc}$, which are now {\bf not} in the loss cone. In the total balance the
number of loss cone stars is only very moderately increased.

In the experiment of measuring the shape of loss cone, we assume the test star can survive only one dynamical time in collisional system, after one orbit it will be ``kicked" to another place in phase space due to interactions with other stars. However, in collisionless limit, if we allow the test star to survive more orbit cycles, test star with much higher $J$ will also have chance to get rid of its angular momentum and be disrupted by BH. Then it is possible that an even larger loss cone region in phase space than what we presented here may exist, and result in a higher TDR. In order to check this, simulations with much more particles are needed and we would like to leave this task for future work.

The orbit type of disrupted stars strongly depends on energy
as we discuss in detail in the previous sections. TD of stars most strongly bound to the BH are
dominated by short-axis tube (SAT) orbits. In intermediate regions saucer orbits dominate, which create a characteristic double peak structure in the last apocenter position of their orbit relative to the equatorial plane. And further out chaotic orbits.

It is known for almost half a century that tidal disruption of stars should occur near SMBH, but only
much more recently the X-ray emission of tidal disruption events has been detected \citep{Komossa2002,KM2008}.
A simple argument on the fallback time for tidal debris by \citet{Rees1988} has led to the prediction of
a characteristic power law of the light curve with time, which can be used to distinguish TD events from
other transients. It is interesting that a SMBH binary can cause characteristic disruptions in such an
otherwise standard TD light curve \citep{LLK2014}. Hayasaki and collaborators claim that
eccentric TD events lead to somewhat longer lived central accretion disks \citep{HSL2013,HSL2015}. It will be very interesting to see whether and how the evolution of tidal debris and the
fallback rate are affected by different orbits of the disrupted stars as discussed here.

It has been claimed that rotation may help to quickly refill loss cones around binary supermassive black holes,
which helps significantly to accelerate shrinking and final coalescence of SMBH binaries in cosmologically
short time scales
\citep{Berczik2006,PBB2011,KHB2013,Khan2014}. In our paper we study by direct $N$-body
simulation the tidal accretion of stars and their orbital parameters in rotating axisymmetric systems.
We confirm the result of \citet{VM2013} that there is an increase in the rate of refilling
of the loss cone, but it is moderate. However, the situation deserves more detailed study, because
a SMBH binary creates a much stronger deviation from spherical symmetry than the one used in our
models with single SMBH. And second the detailed structure of the rotation in a central nuclear
star cluster could affect the enhancement of the loss cone.

\section*{Acknowledgements}

We acknowledge support by Chinese Academy of Sciences through the Silk Road
Project at NAOC, through the Chinese Academy of Sciences Visiting Professorship
for Senior International Scientists, Grant Number 2009S1-5 (RS), and through the
``Qianren" special foreign experts program of China.

The special GPU accelerated supercomputer laohu at the Center of Information and
Computing at National Astronomical Observatories, Chinese Academy of Sciences, funded
by Ministry of Finance of People's Republic of China under the grant ZDY Z2008-2,
has been used for the simulations.

PB acknowledge the special support by the NAS Ukraine under the Main Astronomical
Observatory GPU/GRID computing cluster project.

SZ thank Yohai Meiron for providing the SCF source code which is used in this work.

\newpage

%%%%%%%%%%%%%%%%%%%%%%%%%%%%%%%

\appendix

%%%%%%%%%%%%%%%%%%%%%%%%%%%%%%%

\section{Orbit classification}

%%%%%%%%%%%%%%%%%%%%%%%%%%%%%%%%%%%%%%%%%%      orbit theory: 1
\subsection{Surface of Section}
\label{SoS_Method}

\begin{figure}[htbp]
  \begin{center}
  \includegraphics[width=0.9\columnwidth]{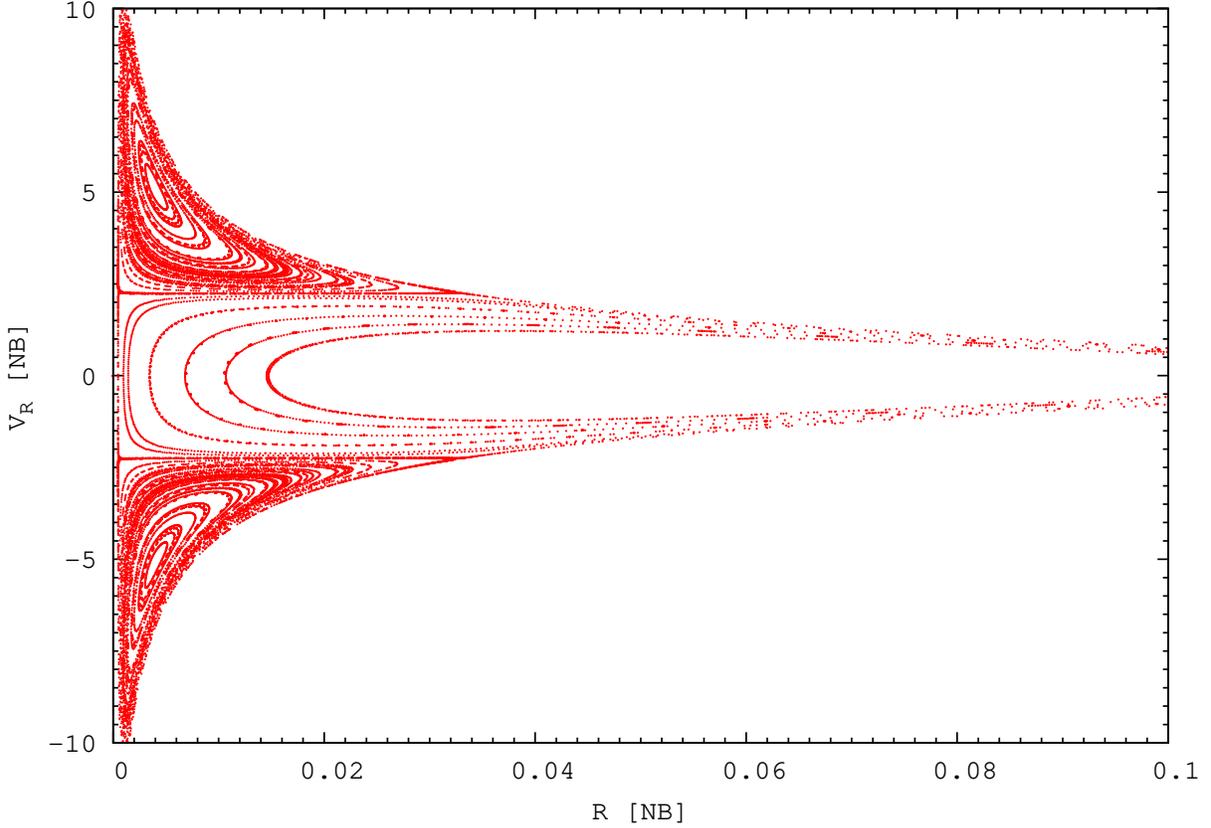}
  \end{center}
  \caption{Surface of section plot. $x$ axis is distance $R$ to origin on equatorial plane. $y$ axis is $v_{R} \equiv dR/dt$ when star go across the equatorial plane.}
  \label{fig_A_SoS}
\end{figure}

From Fig.~\ref{fig_A_SoS} we can see the whole accessible region on $(R,v_R)$ plane is divided into two parts (note that points with opposite $v_R$ actually belongs to same orbit, so this plot is symmetric with horizontal axis). Each part represents a family of orbit. Curves that intersect with $R$-axis are footprints of short axis tube (SAT) orbits, others are of saucer orbits.

%%%%%%%%%%%%%%%%%%%%%%%%%%%%%%%%%%%%%%%%%%      orbit theory: 2
\subsection{Fourier Analysis of $J_x$ evolution}
\label{FA}

\begin{figure}[htbp]
  \begin{center}
  \includegraphics[width=0.47\columnwidth]{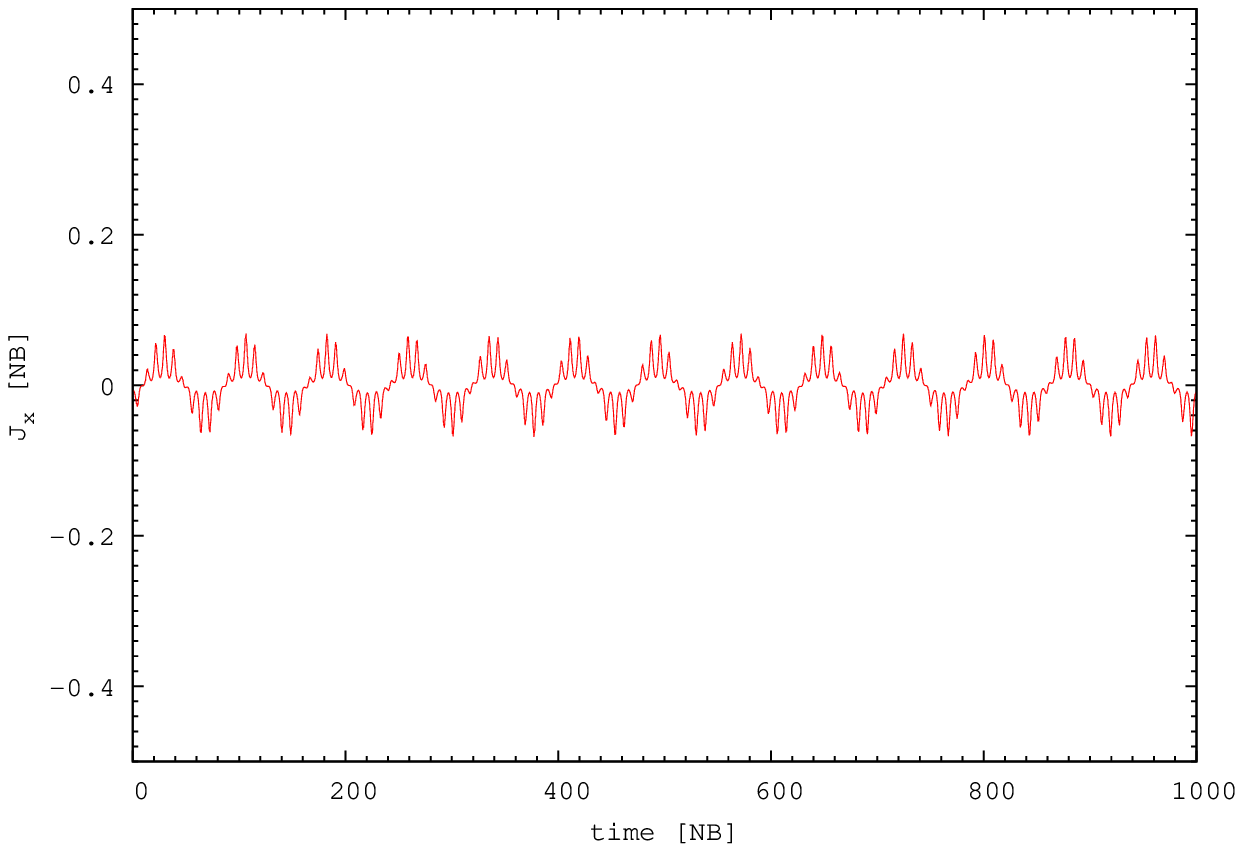}
  \includegraphics[width=0.47\columnwidth]{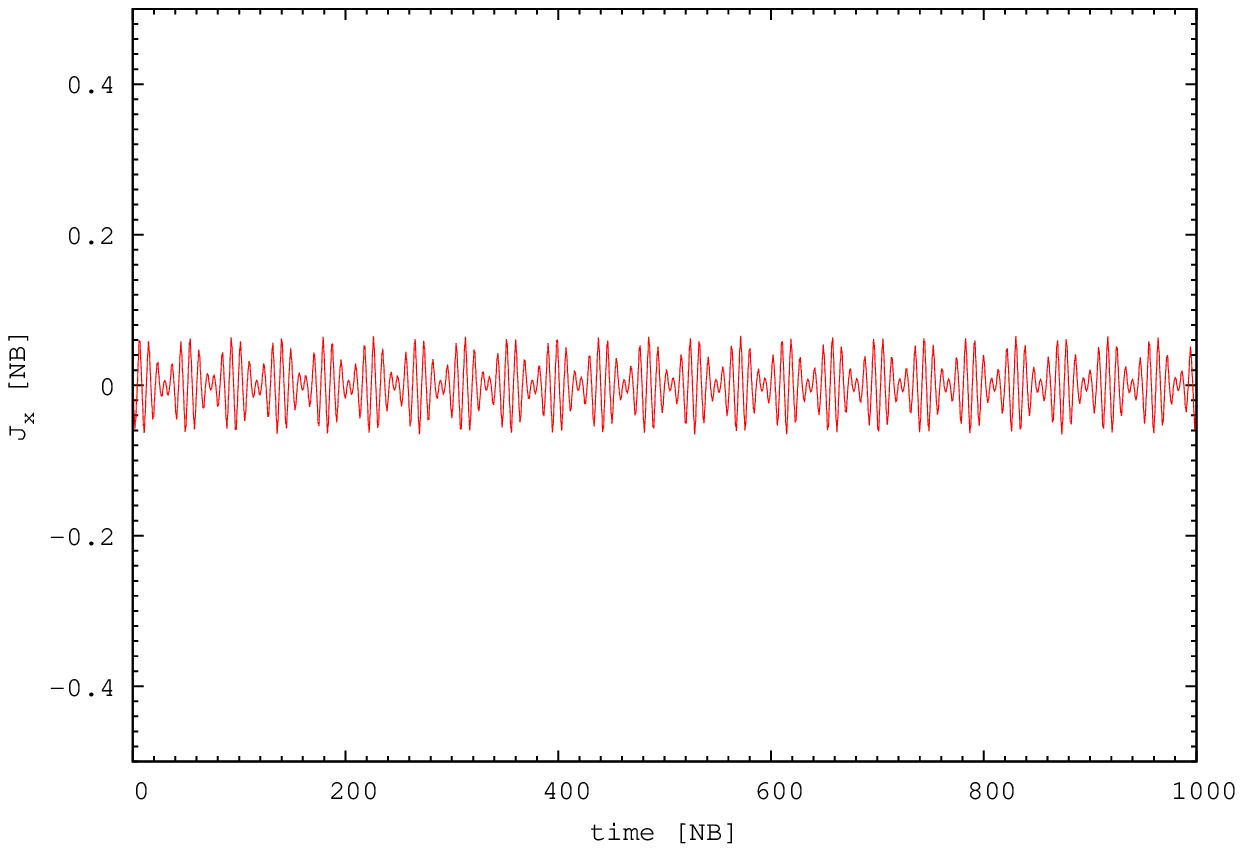}
  \end{center}
  \caption{Time evolution of $J_x$ for SAT orbit (left) and saucer orbit (right).}
  \label{fig_A_J-evo}
\end{figure}

In axisymmetric potential, force is not centripetal hence exerted a torque on the star which will change the $x$ and $y$ components of its angular momentum. Fig \ref{fig_A_J-evo} show the time evolution of $J_x$ for both SAT and saucer orbits. The pattern of $J_x$ and $J_y$ are the same but shifted with a phase of $\pi/2$, so in the following discussion we only focus on $J_x$. Furthermore, the evolution of $J_x$ shows some quasi-periodicity. From eye inspection, one can guess the mathematical expressions for the curves.

\begin{figure}[htbp]
  \begin{center}
  \includegraphics[width=0.47\columnwidth]{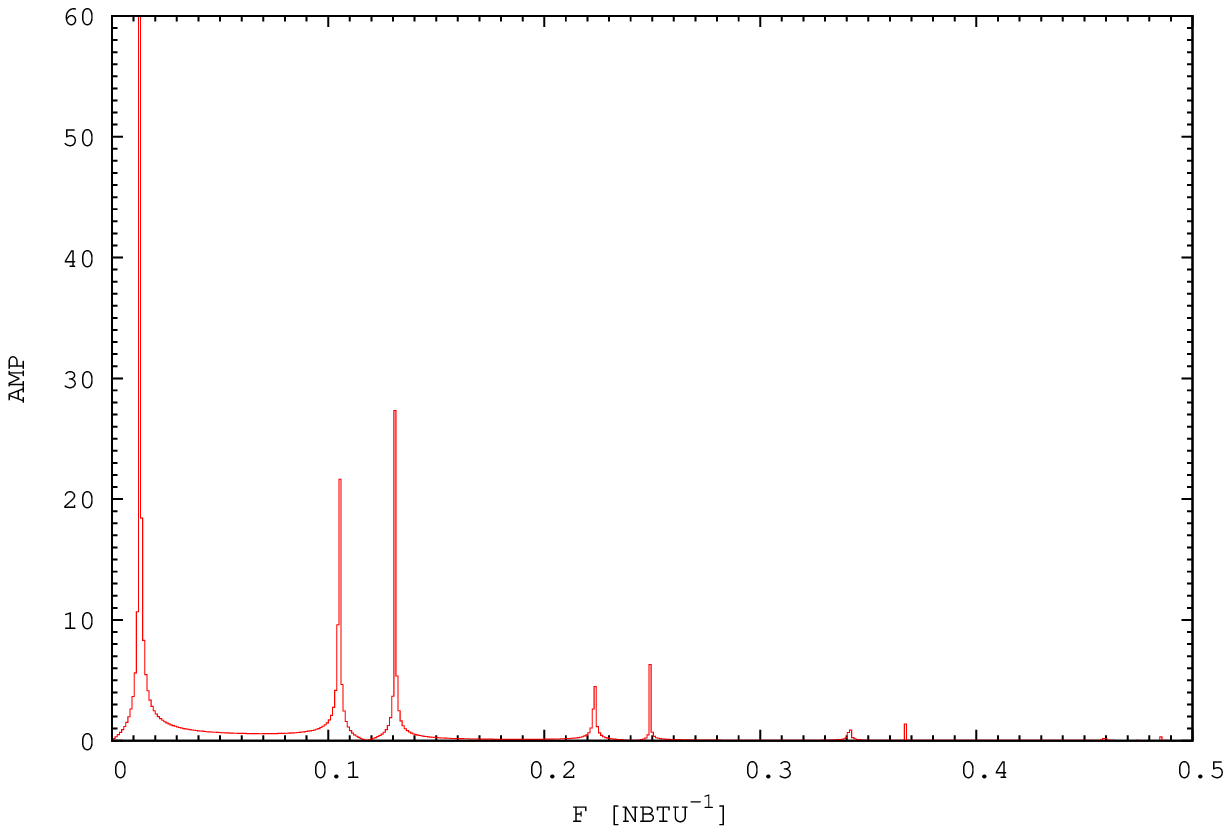}
  \includegraphics[width=0.47\columnwidth]{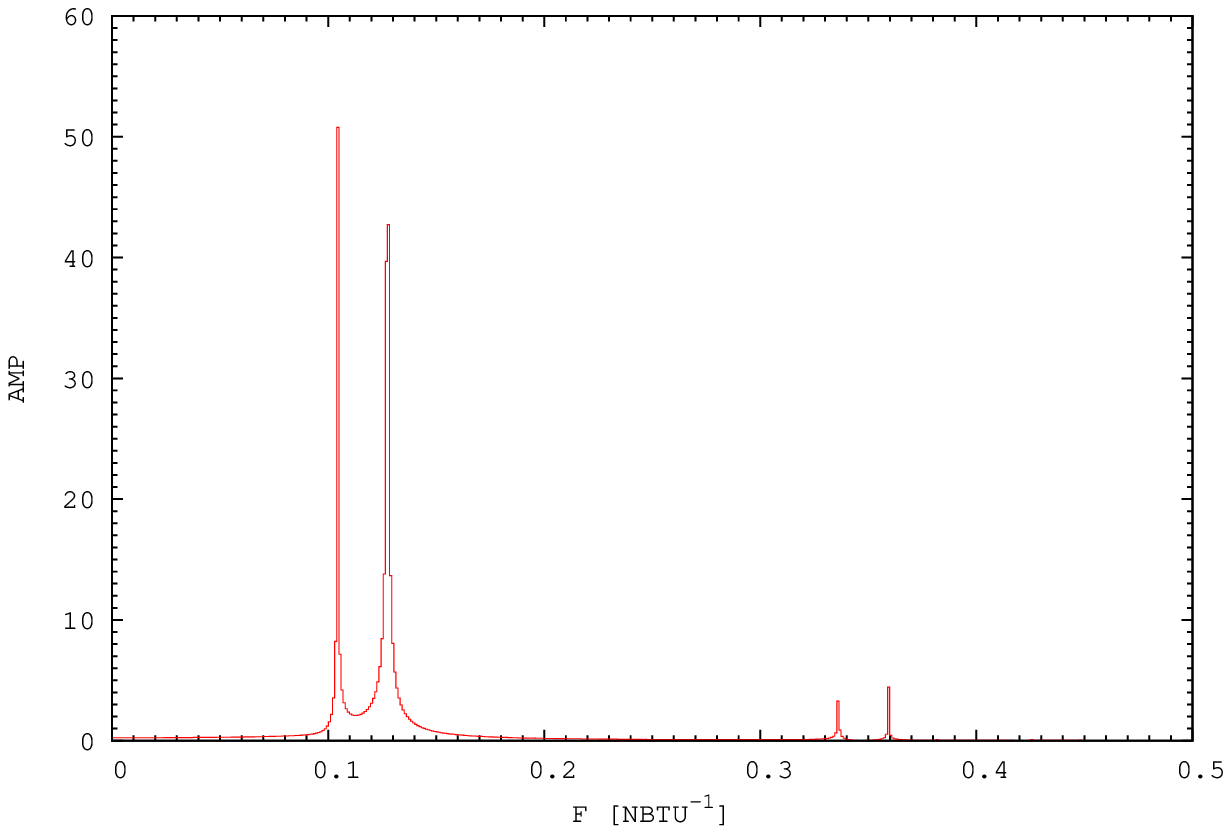}
  \end{center}
  \caption{Fourier frequency distribution of $J_x$. Horizontal axis is frequency in unit of $[T]^{-1}$. Vertical axis is amplitude of corresponding component. Left panel represents SAT orbit, right panel represents saucer orbit.}
  \label{fig_A_freq}
\end{figure}

As shown in Fig.~\ref{fig_A_J-evo}, the curve for SAT orbits seems to be represented by $\rm sin(\rm f_1 t)(1+cos(\rm f_2 t))$ ($\rm f_1 < \rm f_2$), which can be further converted to $\rm sin(f_a t)+\rm sin(f_b t)+\rm sin(f_c t)$ (ignore coefficients before the trigonometric functions), with $\rm f_a = \rm f_1, \rm f_b = \rm f_2 - \rm f_1$ and $\rm f_c = \rm f_2 + \rm f_1$. If we perform a Fourier analysis on this curve, we expect to find 3 principal frequencies: $\rm f_a, \rm f_b, \rm f_c$ in ascending sequence. And these 3 frequencies satisfy the equation $\rm f_a = (\rm f_c - \rm f_b)/2$.

For saucer orbits, the curve seems to be represented by $\rm sin(f_1 t)cos(f_2 t)$ ($\rm f_1 < \rm f_2$), following the same procedure we expected to find 2 principal frequencies: $\rm f_a, \rm f_b$, with $\rm f_a = \rm f_2 - \rm f_1$ and $\rm f_b = \rm f_2 + \rm f_1$.

A demonstration is shown in Fig.~\ref{fig_A_freq}, one can clearly see the 3 principal frequencies for SAT orbits and the 2 for saucer orbits. Some of the small peaks appeared at higher frequencies which is the order harmonics and some are produced from other components.

We use both methods to cross check the validity of orbit classification for the tidally disrupted stars.

%%%%%%%%%%%%%%%%%%%%%%%%%%%%%%%%%%%%%%%%%%%%%%%%%%%%%%%%%%%%%%%%%%%%%%%%%%%%%%%%%%%%%%%%%%%%

\bibliographystyle{apj}
\bibliography{AxisymmetricNuclei}

%%%%%%%%%%%%%%%%%%%%%%%%%%%%%%%%%%%%%%%%%%%%%%%%%%%%%%%%%%%%%%%%%%%%%%%%%%%%%%%%%%%%%%%%%%%%
\end{document}